\documentclass[pra,preprint,showpacs,amsmath,amssymb,nofootinbib%
,eqsecnum%
]{revtex4-1}
\usepackage{graphicx}
\usepackage{mathbbol}
\usepackage{color}
\usepackage[colorlinks=true,urlcolor=blue, linkcolor=blue]{hyperref}

\def\({\left(}
\def\){\right)}

\def\ad{a^\dagger}
\def\a2d{a^{\dagger 2}}
\def\b2d{b^{\dagger 2}}
\def\d#1{#1^\dagger}

\def\eq#1{Eq.~(\ref{eq:#1})}
\def\Eq#1{Equation~(\ref{eq:#1})}
\def\eqs#1#2{Eqs.~(\ref{eq:#1}) \& (\ref{eq:#2})}

\def\fig#1{Fig.\ref{fig:#1}}

\def\Fig#1{Figure~\ref{fig:#1}}

\newcommand\tab[1]{Table~\ref{tab:#1}}
\def\sec#1{Section~\ref{sec:#1}}

\newcommand\ket[1]{\left|\,#1\,\right\rangle}
\newcommand\scalprod[2]{\left\langle\,#1\,\right|\left.#2\,\right\rangle}
\newcommand\om[1]{\omega_{\scriptscriptstyle #1}}

\begin{document}

\title[Continuous-variable quantum computing in the quantum optical frequency comb]{Continuous-variable 
quantum computing in the quantum optical frequency comb}

\author{Olivier Pfister}

\affiliation{Department of Physics, University of Virginia, 382 McCormick Road, Charlottesville, VA 22903, USA}
\email{olivier.pfister@gmail.com}

\begin{abstract}
This topical review introduces the theoretical and experimental advances in continuous-variable (CV) --- i.e.,  qumode-based in lieu of qubit-based --- large-scale, fault-tolerant quantum computing and quantum simulation. An introduction to the physics and mathematics of multipartite entangled CV cluster states is given, and their connection to experimental concepts is delineated. Paths toward fault tolerance are also presented. It is the hope of the author that this review attract more contributors to the field and promote its extension to the promising technology of integrated quantum photonics.
\end{abstract}

\maketitle

\section{Introduction}

\subsection{Quantum computing: revolutionary promise and daunting challenge}

Quantum computing promises exponential speedup for specific tasks~\cite{Montanaro2016}, most notably integer factoring~\cite{Shor1994} and quantum simulation~\cite{Feynman1982}. 
Originally proposed by Feynman, quantum simulation features an exponential speedup over classical computing for the calculation of $N$ quantum systems by solving their Schr\"odinger equation, which entails diagonalizing a $2^{N}\times2^{N}$ Hamiltonian, i.e., an exponential scaling of the complexity of the problem. However, if one were able to exquisitely control, while correcting or staying clear of decoherence, $N$ ``model'' qubits in the laboratory and if one were able to ``dial'' at will interactions between them to implement the precise Hamiltonian to be studied, then the problem would become polynomial in the number of qubits, as these naturally evolve following quantum laws and can be read out using physical measurements (which are included in this polynomial scaling). 

A scalable qubit implementation is therefore  the crux of the power of Feynman's quantum simulator but it is essential to be clear on what scalability means. It is sometimes claimed that simply increasing the dimension of Hilbert space, e.g.\ passing from 2-state qubits to $d$-state qudits, can be useful for quantum simulation. If  the number $N$ of logic units isn't increased in the process, this argument cannot constitute a genuine scalability claim. Indeed, while it is true that the respective Hilbert dimensions verify $d^{N}>2^{N}$ if $d>2$, this constitutes polynomial, not exponential, scaling when $N$ is held constant. As explained above, scaling the quantum computer size with $N$ is what confers a quantum simulator its exponential power, not increasing the Hilbert space of a fixed number of qudits. Hence scalability should always be taken in the sense of an increase of the number of logic units $N$, regardless of their individual Hilbert-space dimension.

Quantum simulation could open the door to scientific discoveries by solving currently intractable problems such as ground state of spin arrays, the Bose-Hubbard model (paving the way to the discovery of room-temperature superconductors), simulating quantum field theory~\cite{Jordan2012,Marshall2015}, and quantum chemistry, from the calculation of energy levels~\cite{AspuruGuzik2005,Huh2015} to optimizing chemical reactions of fundamental societal importance, such as nitrogen fixation~\cite{Reiher2017} and possibly discovering new ones (carbon sequestration). In addition, applications to machine learning have also been  envisaged~\cite{Rebentrost2014,Carleo2017}.

The realization of practical quantum computing, utilizing {\em (i)}, a large enough number of qubits that are, {\em (ii)},  resilient to decoherence, has been known to be a daunting challenge from the inception of the field~\cite{DiVincenzo2001}. Significant progress, however, has been made on the decoherence front, where cutting edge implementations using trapped-ion and superconducting  qubits have now reached the levels of fidelity required for the implementation of quantum error correction\cite{Barends2014,Ofek2016,Gaebler2016}. On the scalability front, quantum control involving arbitrary unitaries has been demonstrated in the 16-dimensional  ground hyperfine manifold of cesium~\cite{Anderson2015}, 2D atomic arrays of 49 atoms were demonstrated~\cite{Saffman2016}, and quantum simulation has been achieved over 51 Rydberg atoms~\cite{Bernien2017} and 53 ions~\cite{Zhang2017}. 

While trapped-ion and superconducting qubits have reached impressive levels of fidelity for quantum gates, they are not yet large-scale quantum systems~\cite{Friis2018}. It has therefore been judiciously proposed to assess the potential of Noisy, Intermediate-Size (50-100 qubit), Quantum (NISQ) computers to achieve quantum advantage, i.e., beyond-classical performance~\cite{Preskill2018}. However, the NISQ concept doesn't capture the whole field. 

\subsection{Quantum computing over continuous variables: a premium on scalability}

Another line of experimental research has placed scalability at the forefront, relying on the remarkable ability of optical parametric oscillators (OPOs) to produce very large numbers of entangled quantum fields~\cite{Pfister2004,Menicucci2008}. Experimental results confirmed this with thousands and a million of entangled quantum modes --- a.k.a.\ {\em qumodes}~\cite{Gottesman2008,Menicucci2008,Furusawa2011,Weedbrook2012} ---  respectively in the frequency~\cite{Pysher2011,Chen2014,Wang2014,Wang2014a} and temporal~\cite{Yokoyama2013,Yoshikawa2016} domains. In this review paper, we focus on the frequency domain implementation in the quantum optical frequency comb (QOFC) of a single OPO, whose scalability mirrors that of its classical counterpart, the OFC of a mode- and carrier-envelope-phase- locked laser,  which emits thousands to millions of classically coherent fields~\cite{Hall2006,Hansch2006}. 

The groundbreaking scalability of the QOFC (as well as of ``temporal combs'' of pulsed quantum fields) comes with the requirement of adopting continuous-variable (CV) quantum information (QI), encoded over dense qumodes rather than over discrete qubits or qudits. Continuous-variable quantum computing (CVQC) was proposed 20 years ago~\cite{Lloyd1999},  offers exponential speedup over classical computing as discrete QC does~\cite{Bartlett2002}, and doesn't suffer from any fundamental impossibility regarding fault tolerance and quantum error correction~\cite{Menicucci2006,Menicucci2014ft}. 

Although quite recent compared to qubit platforms, CVQC has found a natural implementation in the mature field of quantum optics, from the original demonstration of  Einstein-Podolsky-Rosen (EPR)  entanglement~\cite{Einstein1935,Ou1992} to unconditional quantum teleportation~\cite{Furusawa1998}, quantum dense coding~\cite{Li2002}, quantum secret sharing~\cite{Lau2013}, quantum key distribution~\cite{Grosshans2003}, as well as the $\rm C_\text{NOT}$-equivalent $\rm C_\text{SUM}$~\cite{Yoshikawa2008} quantum gate in 2008, 13 years after the first qubit-entangling gates in cavity QED~\cite{Turchette1995} and trapped ions~\cite{Monroe1995}.  

The goal of this review paper is to introduce CVQI to a larger audience already familiar with qubit-based QI, with an eye in particular on the upcoming development of integrated  quantum photonics~\cite{Tanzilli2012,Kues2019} and its extension to the CV domain~\cite{Furst2011,Dutt2015,Lenzini2018,Mondain2019}, which one might expect to enable scalable QI in the same way as integrated electronics enabled scalable classical information. 
 
\section{Continuous-variable quantum computing} 
\subsection{Continuous-variable quantum information: quadrature eigenstates} 

Continuous-variable quantum computing was proposed in 1999 by Lloyd and Braunstein~\cite{Lloyd1999} and the exponential speedup of CVQC, the equivalent of the Gottesman-Knill theorem for qubit-based QC~\cite{Gottesman1999}, was formulated by Bartlett, Sanders, Braunstein, and Nemoto~\cite{Bartlett2002}. Comprehensive reviews of CVQI were written by Braunstein and van Loock~\cite{Braunstein2005a} and by Weedbrook {\em et al.}~\cite{Weedbrook2012}. 

In quantum optical implementations of CVQI, the quantum variables are the ``amplitude'' and ``phase'' quadrature  operators of the quantum electromagnetic field,
\begin{align}
Q &= \frac1{\sqrt2}\(a+\ad\) \\
P &= \frac1{i\sqrt2}\(a-\ad\), 
\end{align}
which are mathematical analogues of the position and momentum observables of a quantum harmonic oscillator of annihilation operator $a$. The term ``quadrature'' refers to the free evolution of the position and momentum of the harmonic oscillator. A correspondence between qubit- and qumode-based quantum information exists and is presented in Table I~\cite{Bartlett2002}. 
\begin{table}[htp]
\label{tab:one}
\begin{center}
\caption{\em Correspondence between qubit- and qumode-based quantum computing~\cite{Bartlett2002}.}
\begin{tabular}{|c|c|}
\multicolumn{2}{c}{}\\
\hline
\textbf{Qubit-based} & \textbf{Qumode-based} \\
\hline
\multicolumn{2}{|c|}{
\em Computational basis\hglue.2\textwidth\ }  \\
\hline
 $\{\ket0,\ket1\}$ & $\{\ket q\}_{q\in\mathbb R}$ \\
$\scalprod k\ell=\delta_{k\ell}$, $k,\ell\in\{0,1\}$ & $\scalprod q{q'}=\delta(q-q')$, $q,q'\in\mathbb R$\\
 $\ket\psi = \psi_{0}\ket0+\psi_{1}\ket1$ &  $\ket\psi = \int dq\,\psi(q)\, \ket q$\\
\hline
\multicolumn{2}{|c|}{
\em Conjugate basis\hglue.2\textwidth\ }  \\
\hline
Hadamard transformed & Fourier transformed\\
$\displaystyle\ket\pm=\displaystyle\frac1{\sqrt{2}}(\ket0\pm\ket1)$ & $\displaystyle\ket p=\frac1{\sqrt{2\pi}}\int e^{ipq}\ket q dq,\quad p\in\mathbb R$ \\
\hline
\multicolumn{2}{|c|}{
\em Single- qubit/qumode  group generators\hglue.2\textwidth\ }\\ 
\hline
Pauli group & Weyl-Heisenberg group of phase-space displacements\\
 $\langle X,Z\rangle$ &  $\langle \{X(\xi)\}_{\xi\in\mathbb R}, \{Z(\varpi)\}_{\varpi\in\mathbb R}\rangle \equiv \langle\{e^{-i\xi P}\}_{\xi\in\mathbb R}, \{e^{i\varpi Q}\}_{\varpi\in\mathbb R}\rangle$ \\
$X\ket j=\ket{j\oplus1},\ j=0,1$& $X(\xi)\ket q=\ket{q+\xi}$\\
$Z\ket j=e^{ij\pi}\ket j,\ j=0,1$& $Z(\varpi)\ket q=e^{i\varpi q}\ket{q}$\\
$X\ket\pm=\pm\ket{\pm}$& $X(\xi)\ket p=e^{-i\xi p}\ket{p}$\\
$Z\ket\pm=\ket\mp$& $Z(\varpi)\ket p=\ket{p+\varpi}$\\
  \hline
\multicolumn{2}{|c|}{
\em Controlled, entangling gates\hglue.2\textwidth\ }\\
\hline
&  $\rm C_{SUM}$:\\
  $\displaystyle{\rm C_X}\ket{j}_{1}\ket{k}_{2}=\ket{j}_{1}\ket{k\oplus j}_{2}$  & $\displaystyle {\rm C_{X}}\ket q_1\ket{q'}_{2}=e^{-i \alpha Q_{1}P_{2}}\ket q_1\ket{q'}_{2}=\ket q_1\ket{q'+ \alpha q}_{2}$ \\
& $\rm C_{PHASE}$:\\
 $\displaystyle{\rm C_Z}\ket{j}_{1}\ket{k}_{2}=e^{i\pi jk}\ket{j}_{1}\ket{k}_{2}$ & $\displaystyle {\rm C_{Z}}\ket q_1\ket{q'}_{2}=e^{i \alpha Q_{1}Q_{2}}\ket q_1\ket{q'}_{2}=e^{i\alpha  qq'}\ket q_1\ket{q'}_{2}$ \\
\hline
\multicolumn{2}{|c|}{
\em Bipartite entanglement\hglue.2\textwidth\ } \\
\hline
Bell state (unnormalized) & EPR state (unnormalizable)\\
\multicolumn{1}{|c}{$\displaystyle\ket{\text B_{00}}_{12}=
\sum_{j=0}^{1}\ket{j}_{1}\ket{j}_{2}$}  & \multicolumn{1}{|c|}{$\displaystyle\ket{\text{EPR}(0,0)}_{12}=\int \ket{q}_{1}\ket{q}_{2} dq=\int\ket{p}_{1}\ket{-p}_{2}dp$}  \\
& \multicolumn{1}{|c|}{$\displaystyle\qquad\qquad\qquad=\sum_{n=0}^{\infty}\ket{n}_{1}\ket{n}_{2}$\ (Schmidt decomp.)}\\
Bell basis & EPR basis\\
\multicolumn{1}{|l}{$\displaystyle\ket{\text B_{k\ell}}_{12}=Z_{1}^{k}X_{1}^{\ell}\ket{\text B_{00}}_{12},\ k,\ell\in\{0,1\}$}  & \multicolumn{1}{|c|}{$\displaystyle \ket{\text{EPR}(\varpi,\xi)}_{12}=Z_{1}(\varpi)X_{1}(\xi)\ket{\text{EPR}(0,0)}_{12},\ \varpi,\xi\in\mathbb R$}\\
\multicolumn{1}{|l}{$\displaystyle\qquad\quad=
\sum_{j=0}^{1}e^{ik\pi(j\oplus\ell)}\ket{j\oplus\ell}_{1}\ket{j}_{2}$} & \multicolumn{1}{|c|}{$\qquad\qquad\ \qquad =
\int e^{i\varpi (q+\xi)}\ket{q+\xi}_{1}\ket q_{2} dq$}  \\
 \hline
\end{tabular}
\end{center}
\vskip -.1 in
\end{table} 
The starting point is the use of the continuous amplitude eigenbasis $\{\ket q\}_{q\in\mathbb R}$ of $Q$ as the computational basis. All discrete sums become integrals and the Hadamard transform over qubits corresponds to the quantum Fourier transform over qumodes. The Pauli group for qubits corresponds to the Weyl-Heisenberg group of all displacements in phase space, generated by quadrature translation operators. Controlled gates and entanglement follow straightforwardly, and have been demonstrated in the laboratory as standalone operations~\cite{Yoshikawa2008,Miwa2009}. Bell bipartite entangled states correspond to Einstein-Podolsky-Rosen (EPR) entangled states~\cite{Einstein1935} (Table I) which are joint eigenstates of the commuting two-mode operators $Q_{1}-Q_{2}$ and $P_{1}+P_{2}$~\cite{Bohr1935} (note that swapping the plus and minus sign in these operators and states is also possible). That means the measurement noise, or quantum standard deviation, for these operators is zero:
\begin{align}
\Delta(Q_{1}-Q_{2})  & = 0    \\
\Delta(P_{1}+P_{2})  & = 0.    
\end{align}
In CVQI lingo, these ``EPR'' operators are also called {\em variance-based entanglement witnesses}~\cite{Hyllus2006} or {\em nullifiers.} As we'll later see, nullifiers  also denote the operatorial logarithms of stabilizers in CV quantum graph theory.

\subsection{Realistic CVQI: squeezed states}

The EPR states have infinite energy, as evidenced by the infinite integrals and sum in Table I, and are therefore unphysical. One can make, however, arbitrarily good approximations of EPR states in the laboratory. These states called two-mode squeezed (TMS) states~\cite{Ou1992}
\begin{equation}
|{TMS}\rangle = \sum_{n=0}^{\infty}\frac{\tanh^nr}{\cosh r}\,|n\rangle_1|n\rangle_2 \ \xrightarrow[r\to\infty]{}\ 2 e^{-r\,}\ket{\text{EPR}(0,0)}_{12}, \label{eq:tms}
\end{equation}
where $r$ is called the squeezing parameter. \Eq{tms} indicates that an EPR state is an infinitely squeezed TMS state. A TMS state can be created directly by an optical parametric amplifier (OPA), e.g.\ a doubly resonant OPO below threshold~\cite{Ou1992,Walls1994}. The corresponding TMS Hamiltonian is
\begin{equation}
H_\text{TMS} = i\hbar\kappa (\d a_{1}\d a_{2}-a_{1}a_{2}),
\end{equation}
where $r=\kappa\tau$ if $\tau$ is the interaction time.  Solving the Heisenberg equations for this Hamiltonian yields 
\begin{align}
Q_{1}(\tau)\mp Q_{2}(\tau)  & = (Q_{1}\mp Q_{2}) e^{\mp r}   \\
P_{1}(\tau)\pm P_{2}(\tau)  & = (P_{1}\pm P_{2}) e^{\mp r}   
\end{align}
which shows clearly that $r\to\infty$ means infinite energy in the total field. For initial vacuum states, one gets the finitely squeezed standard deviations
\begin{align}
\Delta(Q_{1}- Q_{2})  & = e^{-r}    \\
\Delta(P_{1}+ P_{2})  & = e^{-r}.    
\end{align}
We will show in \sec{vbew} that such entanglement witnesses can be generalized to obtain specific signatures of multipartite entanglement from specific multimode squeezing, in particular for cluster entangled states~\cite{Briegel2001,Zhang2006}, which are essential to one-way quantum computing (\sec{owqc}). We now address the universality and fault tolerance of CVQC, and the influence of finite squeezing on the latter.

\subsection{Clifford and Gaussian gates. Exponential speedup and fault tolerance}

\subsubsection{Exponential speedup: the Gottesman-Knill theorem and its CVQC version}

The crucial feature of QC is its exponential speedup over classical computing for specific problems. Such a speedup is present in CVQC as well. The sufficient condition for an exponential speedup is given by the Gottesman-Knill theorem~\cite{Gottesman1999a}, which states that a 
quantum algorithm can be run efficiently on a classical computer if it involves only Clifford gates. Clifford gates, by definition, leave the Pauli group globally invariant. This means that the transform $\mathcal P'$ of a Pauli operator $\mathcal P$ by a Clifford operator $\mathcal C$, 
\begin{equation}
\mathcal P' = \mathcal C \mathcal P \mathcal C^{-1},
\end{equation}
is also a Pauli operator. Gottesman showed that Clifford-gate algorithms over $N$ qubits can be modeled classically in the Heisenberg picture by tracking the evolution of only $N$ operators~\cite{Gottesman1999a}, which is exponentially efficient since the Hilbert space of quantum states is of dimension $2^{N}$. However, this approach fails when the quantum algorithm contains at least one non-Clifford gate, which requires in principle  that one consider the full $2^{N}$ dimensionality of the Hilbert space for the classical simulation of the quantum algorithm. In that case, Feynman's argument for QC (that the needed resource is only $N$ qubits versus $2^{N}$ classical bits) is in full force and yields an exponential speedup for QC. An example of a non-Clifford gate is the $\pi/4$ rotation around $\hat z$ which transforms Pauli operator $X$ into $(X+Y)/\sqrt2$, which isn't a Pauli operator since the Pauli group is a multiplicative one. The $\pi/4$ rotation is present in the quantum Fourier transform in Shor's algorithm, for example~\cite{Nielsen2000}.

For CVQC, one must then seek the group of transformations that leave the Weyl-Heisenberg group globally invariant (i.e., the normalizer of the Weyl-Heisenberg group in mathematical terms). As Bartlett {\em et al.}\ first established~\cite{Bartlett2002}, the group that normalizes displacements is that of all Gaussian operations, i.e., the group of unitary evolution operators corresponding to Hamiltonians  at most of quadratic order in the quantum fields (i.e., in creation/annihilation operators or quadrature operators). Such evolution operators possess Gaussian Wigner functions and, when operating on a quantum state of a Gaussian Wigner function, leave its Gaussian character unchanged, even though  the Wigner function of the evolved quantum state may change through displacement, rotation, squeezing or shearing. 

Hence, non-Gaussian operations are crucial to ensure that CVQC yields an exponential speedup. Such operations correspond to quantum evolution under a Hamiltonian that's of cubic order (or higher) in the fields. Cubic Hamiltonians were also proven to be the minimum-order necessary resource to enable the generation of Hamiltonians of arbitrary order in Lloyd and Braunstein's first proposal of CVQC~\cite{Lloyd1999}. 

Other avenues, more feasible in the laboratory, also exist, such as projection in the Fock basis via photon-number-resolved (PNR) detection~\cite{Lita2008}, photon subtraction~\cite{Wenger2004,Ourjoumtsev2006}, or photon addition~\cite{Zavatta2004}. For example, Gottesman, Kitaev, and Preskill proposed the use of PNR detection as the sole non-Gaussian resource needed to implement cubic phase gate $\exp(i\gamma Q^3)$~\cite{Gottesman2001}.  

\subsubsection{Fault tolerance}

To the best of current knowledge, non-Gaussian operations are critical to fault tolerance of CVQC. It is a somewhat counterintuitive feature of CVQC that, while the Clifford-Gaussian correspondence holds when considering QC's exponential speedup, it does not hold for other concepts such as  Bell inequality violation~\cite{Bell1987}, entanglement distillation~\cite{Eisert2002}, and quantum error correction~\cite{Niset2009}: all these operations can  be implemented over qubits using Clifford resources, but cannot be implemented over qumodes using solely Gaussian resources. 

However, the use of non-Gaussian resources, such as Fock states, PNR measurements, or field-cubic Hamiltonians~\cite{Lloyd1999} remedies the situation and removes all impossibilities. 

Fault tolerance and quantum error correction deserve a bit more detail. The effect of finite squeezing on fault tolerance is an important question, as one might be tempted to argue that the intrinsically ``fuzzy'' wavefunctions $\psi(q)$ used in CVQI will unavoidably lead to non-correctible errors, as is the case for classical analog computing. The issue is, while it is deemed straightforward to distinguish the orthogonal $\ket0$ and $\ket1$ states of a qubit, it is impossible to distinguish the orthogonal $\ket q$ from $\ket{q+\delta q}$ if $\delta q\ll\Delta Q_{\psi}$, even in a squeezed state $\Delta Q_{\psi}\propto \exp(-r)$,  $\Delta Q_{\psi}$ being the standard deviation of $Q$ in state $\ket\psi$.

As an example, one can ask how many CV teleportation~\cite{Furusawa1998} steps can be concatenated before the teleportation fidelity reaches the classical limit of 50\%.\footnote{This is an especially meaningful question since the teleportation gate can be used as a primitive for QC~\cite{Gottesman1999}.} The answer is $n_\text{max}=\exp(2r)$ if $r$ is the squeezing parameter of the TMS states used as teleportation channels~\cite{Yonezawa2007}. This means $n_\text{max}=2$ for 3 dB of squeezing and 10 for 10 dB of squeezing. Other studies of uncorrected CVQC made clear that quantum error correction is needed~\cite{Ohliger2010,Ohliger2012} (but this is equally true of qubit-based platforms). 

While it is not known whether it might be possible to directly correct CV errors,  a path to fault-tolerant CVQC does exist: Gottesman, Kitaev, and Preskill (GKP) proposed the discrete encoding of a qubit in an oscillator to address  small CV drifts~\cite{Gottesman2001}, and this encoding was applied to CVQC by Menicucci to prove the existence of a fault tolerance threshold~\cite{Menicucci2014ft}. Therefore, infinite squeezing is {\em not} a requirement for fault-tolerant CVQC. As we will mention later,  the amount of squeezing required for fault tolerance is actually not unreasonable. The GKP error encoding relies on the creation of GKP resource states that are comb-like in quadrature quantum phase space. Experimental realization of GKP states is an arduous endevor. Proposals have been made for the generation of optical GKP states~\cite{Vasconcelos2010,Motes2017,Weigand2018,Eaton2019} and an experimental realization over the phononic vibration field of trapped ion was recently performed~\cite{Fluhmann2019}.

Finally, interesting results have been obtained recently on implementing QC over {\em qubits} using a GKP encoding: in that situation, Baragiola et al.\ have shown that Gaussian operations are enough, along with the GKP encoding, to achieve universal, fault-tolerant QC with no cubic-phase gate needed~\cite{Baragiola2019}. Also intriguing are recent investigations of subuniversal  quantum computing {\em \`a la} boson sampling~\cite{Aaronson2010}, such as Gaussian boson sampling~\cite{Hamilton2017} and CV instantaneous quantum computing~\cite{Douce2017}, which have both been proven to be hard classically.

\section{Measurement-based, one-way quantum computing}\label{sec:owqc}
\subsection{Introduction. Cluster states}

An equivalent (but still universal) alternative to the circuit model of universal QC~\cite{Nielsen2000} is that of measurement-based QC~\cite{Gottesman1999} and, in particular, one-way QC, based on cluster entangled states~\cite{Briegel2001}. One-way QC starts from a cluster state or ``quantum computing substrate,'' an entangled qubit lattice which contains all the entanglement that can ever be needed by a quantum algorithm. Quantum computing can then proceed solely by single-qubit measurements which inform feed-forward unitaries on the lattice neighbors~\cite{Raussendorf2001}.

From this description, it is clear that the concept of cluster state is a central one. Indeed, while the cluster state clearly must be a multipartite entangled state, it cannot be just any multipartite state. It is well known that multipartite entanglement differs fundamentally from bipartite entanglement in that there exists distinct families of LOCC-equivalent entangled states, where LOCC stands for local operations and classical communication. For example, the W states, e.g.\ $\ket{001}+\ket{010}+\ket{100}$, and the  Greenberger-Horne-Zeilinger (GHZ) states~\cite{Greenberger1989}, e.g.\ $\ket{000}+\ket{111}$, are not LOCC equivalent~\cite{Dur2000}. Cluster states are not LOCC equivalent to either GHZ or W states for 4 qubits and more.
 
A cluster state is canonically defined as qubits initialized in the $\ket+$ state and interacting via $\rm C_{Z}$ gates in a 2D pattern, typically a square lattice (although other lattices are also possible)~\cite{Briegel2001}. It is convenient to represent cluster states as graphs, see \fig c. We'll call such graphs ``canonical'' graphs throughout the paper --- as opposed to the CV $\mathcal H$ graphs that we'll define later.
\begin{figure}[htb]
\centerline{\includegraphics[width=.3\columnwidth]{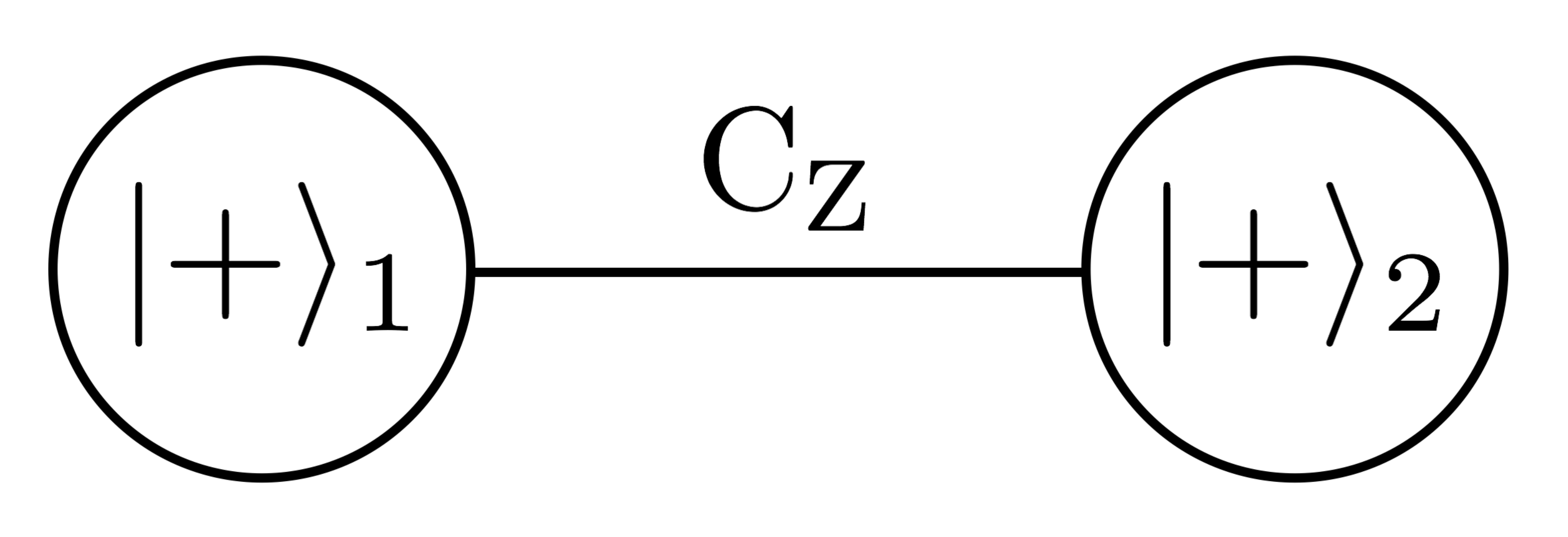}}
\caption{Graphical representation of a cluster state: vertices denote qubits in the $\ket+$ state, edges denote $\rm C_{Z}$ gates.}
\label{fig:c}
\end{figure}
The effect of measurements on a cluster can be understood easily as a multipartite generalization of teleportation: in regular teleportation, a bipartite entangled state (LOCC-equivalent to a cluster state) is the quantum resource and Alice's choice of measurement basis decides the quantum gate applied to the teleported state~\cite{Jozsa2005}. In order to realize the universal QC gate set, the cluster state must be a 2D lattice, such as a square lattice, so as to allow for two-qubit gates~\cite{Raussendorf2001,Jozsa2005,Briegel2009}. 

\begin{figure}[htb]
%
\centerline{\includegraphics[width=\columnwidth]{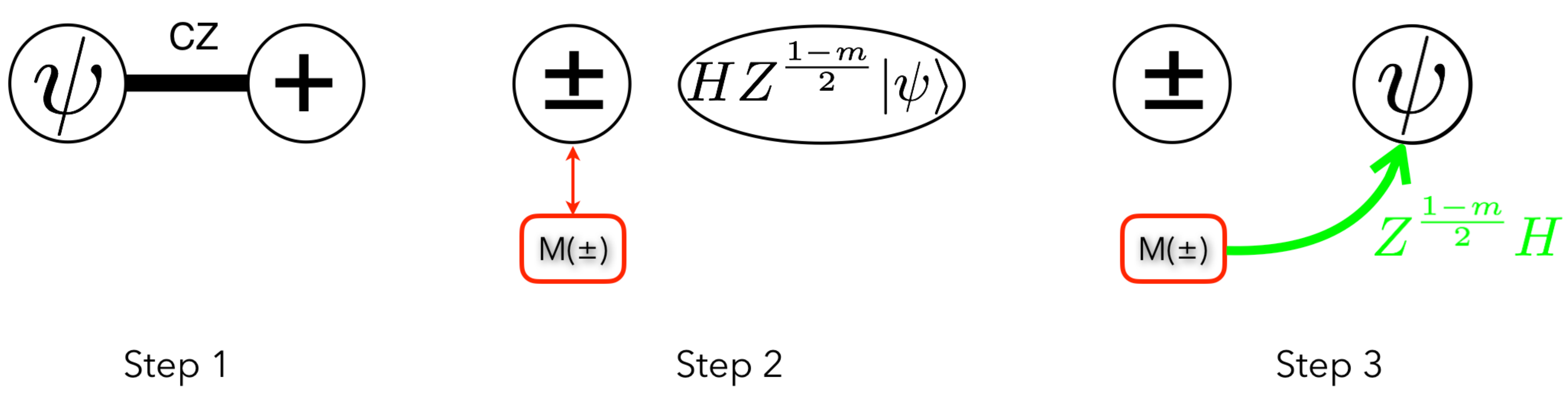}}
\caption{
Quantum teleportation using a cluster-state. Step 1: the state to be teleported is fused to the simplest possible cluster state. Step 2: a measurement is made in the $\ket\pm$ basis, projecting the neighbor qubit. Step 3:  the (random) measurement result $m=\pm1$ is used to feed forward onto the neighbor qubit, thereby deterministically placing it into the state to be teleported.}
\label{fig:qt}
\end{figure}

A simple illustration of measurement-based quantum processing is given by the teleportation gate sequence of \fig{qt}.

The cluster state concept is at the heart of QC scalability as featured in this paper: if one is able to generate the whole cluster state in a scalable manner, then all QC needs to proceed is single-qubit measurements and feedforward. This is the CVQC approach this paper discusses in the QOFC. We will also mention implementations in the time domain which are very similar is spirit. 

\subsection{The CV cluster state}

One-way CVQC has been formulated using CV cluster states~\cite{Menicucci2006,Gu2009}. Using Table I, we can easily see how to generate CV cluster states from the qubit definition~\cite{Zhang2006}: they are created by applying $\rm C_{Z}$ (or $\rm C_\text{PHASE}$) gates along a square lattice of qumodes in phase-quadrature eigenstates $\ket{p=0}$. For illustration, we can compare the literal expression of a two-qubit cluster state,
\begin{equation}
{\rm C_{Z}}\ket+_{1}\ket+_{2} = \ket0_1\ket0_{2}+\ket0_1\ket1_{2}+\ket1_1\ket0_{2}-\ket1_1\ket1_{2},\label{eq:2qb}
\end{equation}
to that of a two-qumode cluster state
\begin{equation}
e^{i\alpha Q_{1}Q_{2}}\ket{p=0}_{1}\ket{p'=0}_{2} = \frac1{2\pi}\iint dq\,dq'\,e^{i\alpha qq'}\,\ket q_{1}\ket{q'}_{2}.\label{eq:2qm}
\end{equation}
Again, these cluster states are infinitely squeezed, thus unphysical; in the laboratory, we can only create finitely phase-squeezed states, which are created by single-mode squeezers, e.g.\ degenerate OPAs, and then apply the $\rm C_\text{PHASE}$, a.k.a.\ quantum nondemolition (in the backaction evading sense~\cite{Braginsky,Caves1980a,Yurke1985}), or spring coupling,  gates~\cite{Miwa2009}. 

\subsection{Bottom up scalability}

This canonical method of building an $N$-qumode cluster state would therefore require $N$ degenerate OPAs for the initial states and a couple of OPAs per entangling gate. While this method scales  linearly with such experimental resources, it is nonetheless not the most efficient way to generate a CV cluster state. A first improvement consists in noticing that the $N$-mode cluster state is a Gaussian resource whose generation protocol can be re-cast as a Bloch-Messiah decomposition~\cite{Braunstein2005} consisting in $N$ single-mode squeezers ``sandwiched'' between two $N$-mode interferometers. When the input states are vacuum ones, the first interferometer is irrelevant and any such Gaussian state can be created from $N$ single-mode squeezers followed by one $N$-mode interferometer, which greatly simplifies the protocol by replacing all nonlinear optical $\rm C_\text{PHASE}$ gates with linear optical interferometers~\cite{vanLoock2007,Yukawa2008a}. One can, however, find even more compact methods for generating cluster states.

\subsection{Top-down scalability: CV cluster state generation in the QOFC}

In the above protocols, each single-mode squeezer is a degenerate OPA that is essentially an OPO cavity with a multitude of resonant modes, all of them but one unused! Instead of scaling to $N$ qumodes using $N$ OPO's, it is then possible (and experimentally more tractable) to use the whole QOFC of a single OPO. The first idea to use $N$-mode squeezing to generate $N$-mode entanglement involved only GHZ states~\cite{Pfister2004}. It was then extended to proposing cluster-state generation~\cite{Menicucci2007} and, finally, proposing the generation of a square cluster lattice in a single OPO~\cite{Menicucci2008}. Experimental realizations followed in the QOFC~\cite{Pysher2011,Chen2014,Roslund2014} as well as in the pulsed ``temporal comb'' regime~\cite{Yokoyama2013,Yoshikawa2016}. We now detail the mathematical formalism used to describe CV cluster states, as it also informs the methods for their generation, before describing experimental implementations of large-scale CV cluster states in the QOFC.

\section{Graph states}\label{sec:vbew}

\subsection{Canonical graph states, stabilizers, and nullifiers}

\subsubsection{Qubits}

As we mentioned earlier, qumode cluster states can be represented as canonical graphs whose vertices are phase-squeezed states and edges are $\rm C_\text{PHASE}$ gates. These are directly deduced from the qubit formalism using the correspondence of Table I~\cite{Bartlett2002}.  An important feature of any graph state $\ket\psi$ (over qubits, qudits, or qumodes) is that it is a stabilizer state, i.e., is uniquely defined by a group $\mathcal S$ of operators $S$ that leave $\ket\psi$ invariant:
\begin{equation}
\forall S\in\mathcal S, \ S\ket\psi=\ket\psi.
\end{equation}
For qubit cluster states, the multiplicative stabilizer group $\mathcal S$ is generated by all possible products of its generators, which are constructed by taking the Pauli $X$ operator on any vertex and the Pauli $Z$ operator on all its graph neighbors:
\begin{equation}
\mathcal S = \left\langle X_{j}\bigotimes_{k\in\mathcal N_{j}} Z_{k}\right\rangle_{j \text{ spans }\mathcal S}, \label{eq:g}
\end{equation}
where $\mathcal N_{j}$ denotes the neighborhood  (i.e., the set of all edge-connected qubits) of qubit $j$. Again for illustration purposes, applying this to the simple cluster state of \eq{2qb} yields
\begin{align}
(X_{1}Z_{2}){\rm C_{Z}}\ket+_{1}\ket+_{2} &= \ket1_1\ket0_{2}+\ket1_1(-\ket1_{2})+\ket0_1\ket0_{2}-\ket0_1(-\ket1_{2})\\
&={\rm C_{Z}}\ket+_{1}\ket+_{2}.
\end{align}
These generators of the stabilizer group thus provide a Heisenberg picture of a cluster state (akin to that used in Gottesman's treatment of Clifford quantum algorithms~\cite{Gottesman1999a}) and constitute an efficient prescription for what observables to measure in the laboratory in order to certify that a cluster state was made. As mentioned above, we call these observables variance-based entanglement witnesses: if all generators of \eq g are measured with no quantum noise, using many copies of $\ket\psi$, then $\ket\psi$ is an eigenstate of the generators, i.e., a cluster state whose graph can be reconstructed from the edge structure of the neighborhoods. We now translate this to CV~\cite{vanLoock2007,Menicucci2011}. 

\subsubsection{Qumodes}

As per Table I~\cite{Bartlett2002}, we have the correspondence
\begin{align}
X & \to X(\xi) = e^{-i\xi P}  \\
Z & \to Z(\varpi) = e^{i\varpi Q}.  
\end{align}
The stabilizer group generators, which were both unitary and Hermitian for qubits, are now only unitary for qumodes and can be expressed as 
\begin{equation}
X_{j}(\xi_{j})\bigotimes_{k\in\mathcal N_{j}} Z_{k}(\varpi_{k}) = \exp\left\{-i\xi_{j}\left[P_{j}-\sum_{k\in\mathcal N_{j}}\frac{\varpi_{k}}{\xi_{j}}Q_{k}\right]\right\}.\label{eq:cvs}
\end{equation}
Its action on example state of \eq{2qm} gives
\begin{align}
[X_{1}(\xi) Z_{2}(\varpi)]{\rm C_{Z}(\alpha)}\ket{p=0}_{1}\ket{p'=0}_{2} & = e^{-i\xi P_{1}}e^{i\varpi Q_{2}}\frac1{2\pi}\iint dq\,dq'\,e^{i\alpha qq'}\,\ket q_{1}\ket{q'}_{2}   \\
& = \frac1{2\pi}\iint dq\,dq'\,e^{i\alpha qq'+i\varpi q'}\,\ket {q+\xi}_{1}\ket{q'}_{2}   \\
& = \frac1{2\pi}\iint dq\,dq'\,e^{i\alpha qq'}e^{iq'(\varpi-\alpha\xi)}\,\ket {q}_{1}\ket{q'}_{2}   
\end{align}
which yields stabilization for $\varpi/\xi=\alpha$. (Note that qumode stabilizer states, unlike qubit ones, can have weighted-edge graphs.)

As before, we will be interested in the variance-based entanglement witnesses. These will be given by the Hermitian operators that were exponentiated to give the stabilizers of a states $\ket\psi$ as per \eq{cvs}. For stabilizers to have an eigenvalue of 1, these Hermitian operators must have a zero eigenvalue and will be called {\em nullifiers} for this reason. Nullifiers are quadrature operators and can be directly measured using standard quantum optics tools such as balanced homodyne detection and RF networks (splitters/combiners and phase shifters). Defining the vectors $\vec Q=(Q_{1},\dots,Q_{N})^{T}$ and $\vec P=(P_{1},\dots,P_{N})^{T}$ for $N$ qumodes, we can then write the vector equation
\begin{equation}
\(\vec P - {\bf V} \vec Q\)\ket\psi = \vec 0\ket\psi.
\end{equation}
where {\bf V} is the mathematical adjacency matrix of the cluster graph, whose entries ${\bf V}_{ij}$ are nonzero if and only if there exits an edge between vertices $i$ and $j$. 

\subsection{$\mathcal H$(amiltonian) graph states and their connection to canonical graph states}

A different type of graph, the $\mathcal H$(amiltonian) graph, can also be defined. It is highly relevant experimentally and is relatable to {\bf V}. The idea of the $\mathcal H$ graph stems from the proposal to generate multipartite entanglement using multimode squeezing of Hamiltonian~\cite{Pfister2004,Bradley2005}
\begin{equation}
H= i\hbar\kappa\sum_{i<j}G_{ij} (\d a_{i}\d a_{j} - a_{i} a_{j}) \label{eq:H}
\end{equation}
It is easy to show that this Hamiltonian yields the system of equations of motion
\begin{equation}
\frac {d\vec Q}{dt} = \kappa\,{\bf G}\,\vec Q,
\end{equation}
where {\bf G} is the matrix of entries $G_{ij}$ and is the adjacency matrix of the $\mathcal H$ graph. Diagonalizing {\bf G} provides the squeezing parameters (eigenvalues) and the squeezed multimode observables (eigenvectors). We will always assume the initial state is the vacuum. See \fig g.
\begin{figure}[htb]
\centerline{\includegraphics[width=.75\columnwidth]{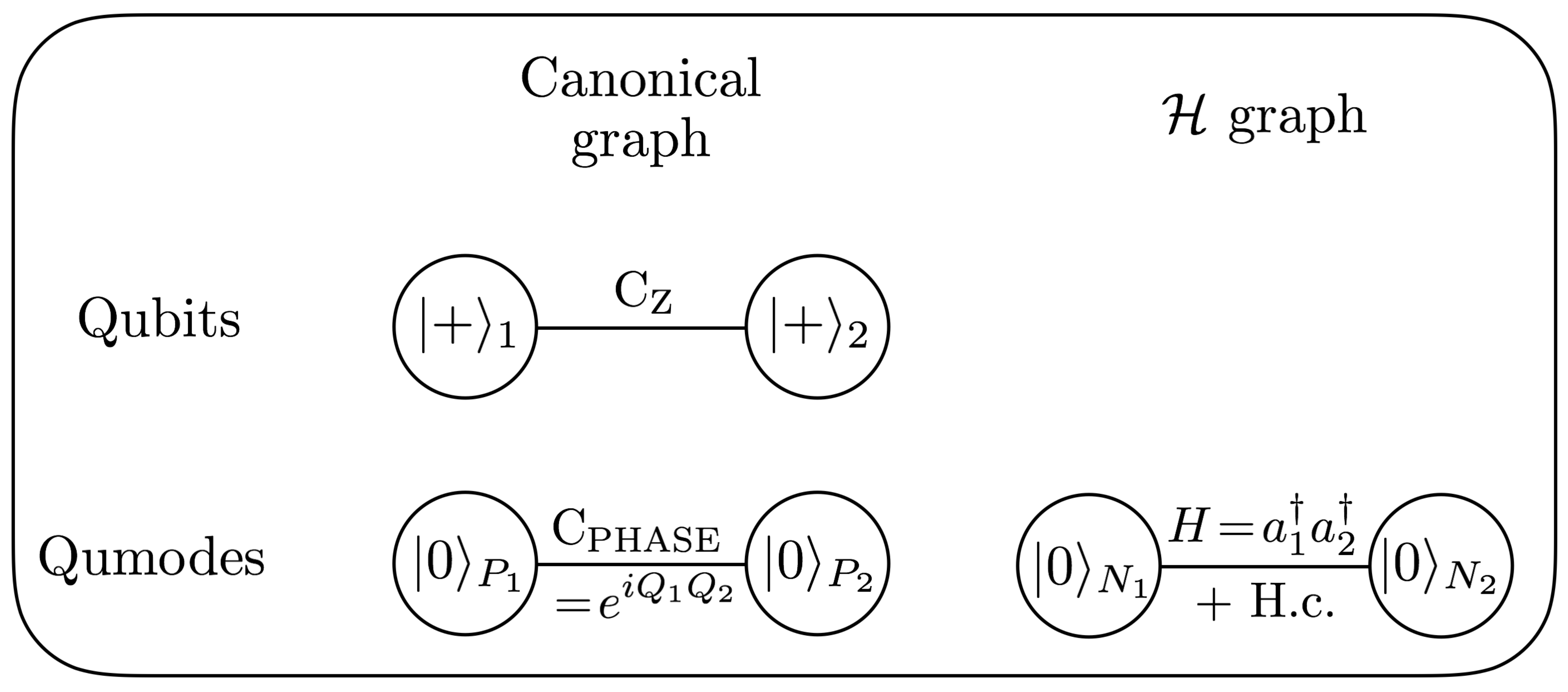}}
\caption{The different quantum graph types. Note that the starting states are vacuum states in the $\mathcal H$-graph case, as $N_{j}=\d a_{j}a_{j}$.}
\label{fig:g}
\end{figure}
The question of the relation of the $\mathcal H$ graph to the canonical graph or, equivalently, of {\bf G} to {\bf V}, isn't an easy one but it has been resolved for infinite~\cite{Menicucci2007} and finite~\cite{Menicucci2011} squeezing. An interesting theorem is~\cite{Zaidi2008}
\begin{equation}\label{eq:theorem}
{\bf G}={\bf G}^{-1} \Rightarrow {\bf G}={\bf V}.
\end{equation}

\subsection{An example: the GHZ graph}

\begin{figure}[htb]
\centerline{\includegraphics[width=.5\columnwidth]{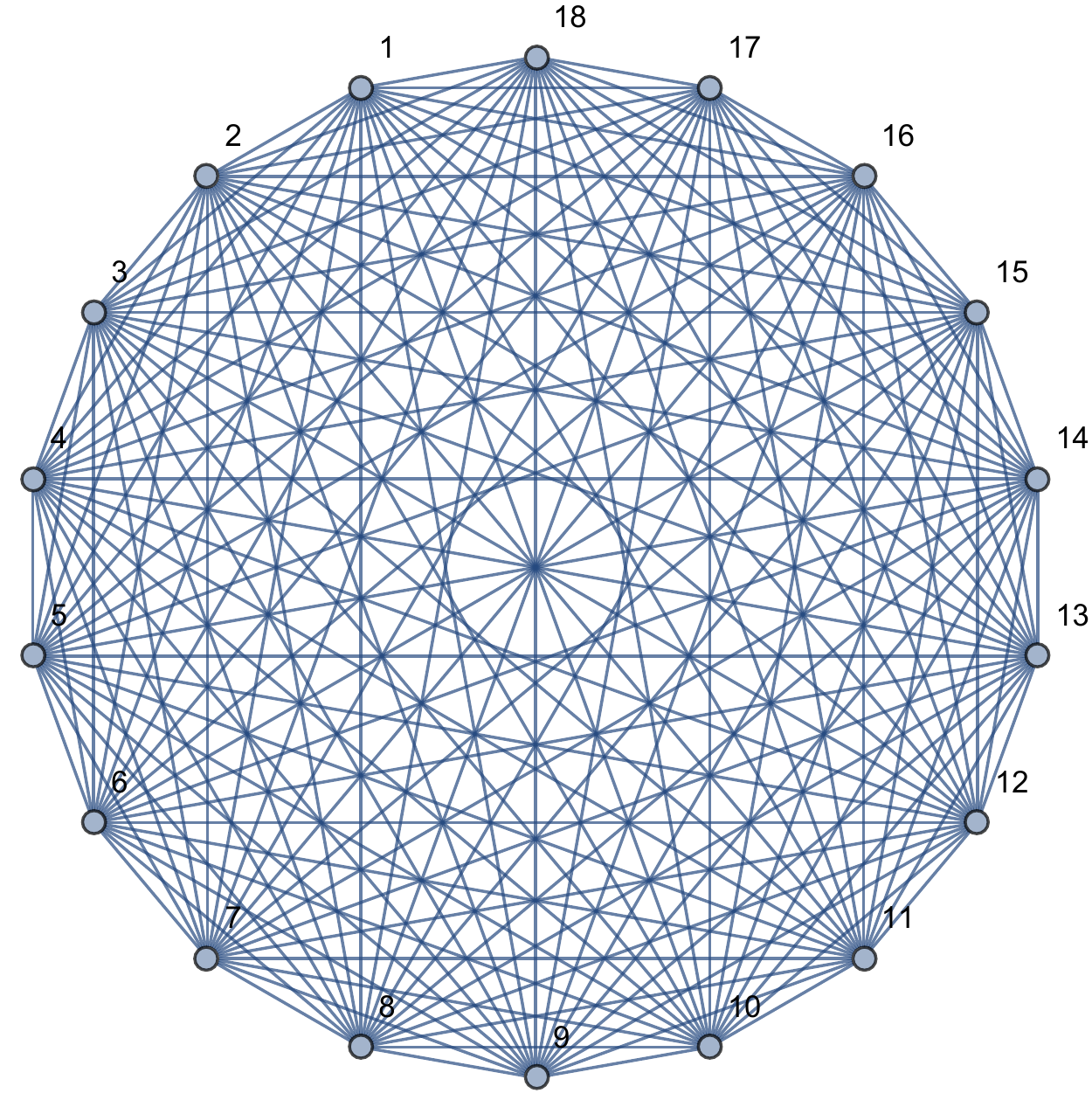}\includegraphics[width=.5\columnwidth]{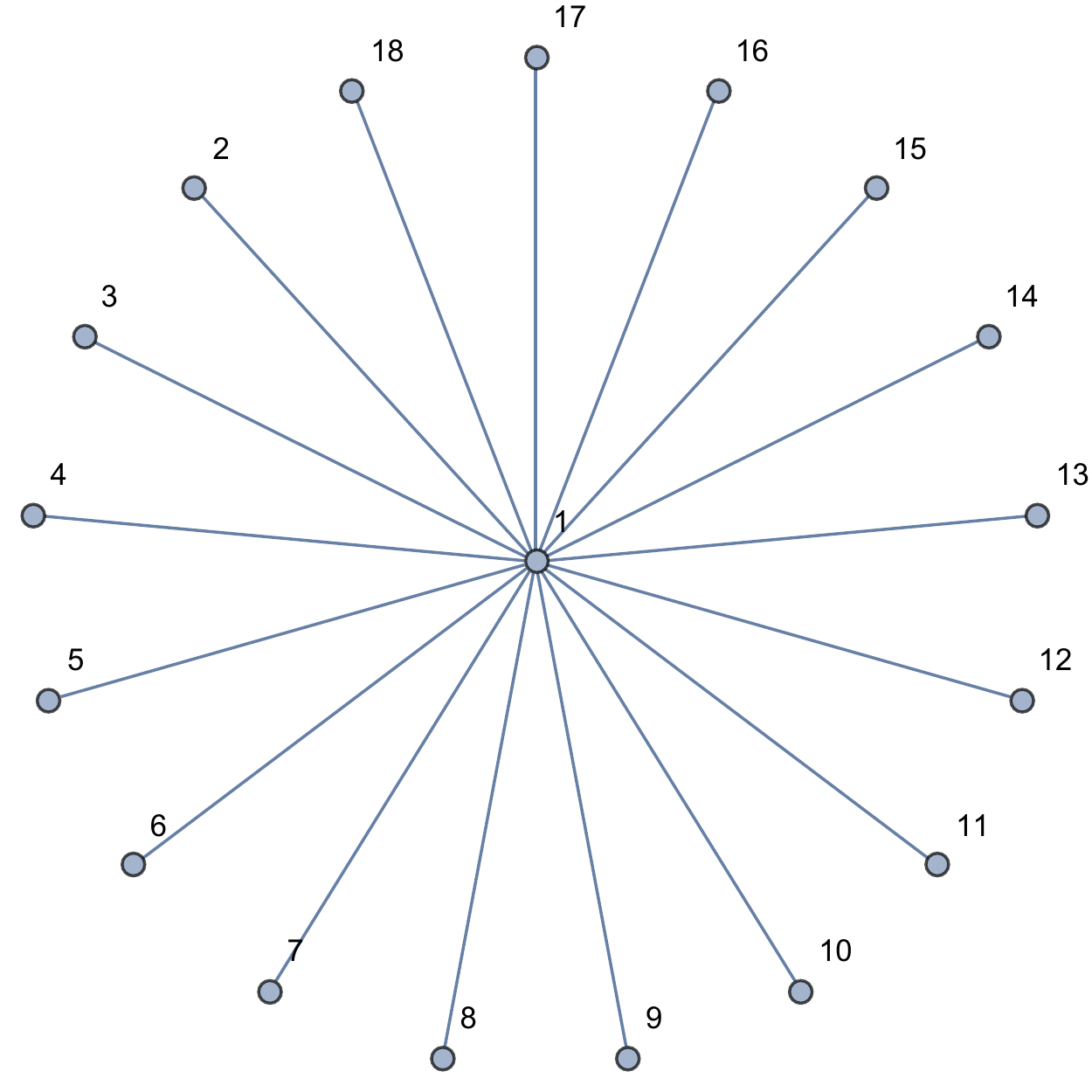}}
\caption{Left, the $\mathcal H$ graph of a GHZ state \eq{ghz} for $N=18$. Right, the corresponding canonical graph obtained from the nullifiers of \eqs{n1}{n2}.}
\label{fig:ghz}
\end{figure}

The first proposal of compact generation of multipartite entanglement in the QOFC~\cite{Pfister2004} was for a GHZ state. The Hamiltonian in this case has all possible TMS terms, which leads to the  following {\bf G} matrix
\begin{equation}\label{eq:ghz}
{\bf G} = \kappa\begin{pmatrix}
0 & 1 & \cdots & 1 \\
1 & \ddots & \ddots & \vdots \\
\vdots & \ddots & \ddots & 1 \\
1 & \cdots & 1 & 0
\end{pmatrix}.
\end{equation}
This {\bf G} matrix corresponds to the complete graph, an example of which is shown in \fig{ghz}, left. Note that {\bf G} is not self-inverse in this case.
Solving the Heisenberg equations~\cite{Pfister2004} yields the nullifiers 
\begin{align}
P_{+} &= e^{-(N-1)r}\,\sum_{j=1}^{N}P_{j} \label{eq:Nr}\\
Q_{jk} & = e^{-r}(Q_{j}-Q_{k}),\quad \forall j\neq k.
\end{align}
Note the remarkable squeezing boost by the mode number in \eq{Nr}~\cite{Pfister2004}. In the limit of infinite squeezing, these nullifiers are those of the following GHZ state~\cite{Braunstein2003a}:
\begin{equation}
\ket{\text{GHZ}} = \int dq\,\ket q_{1}\ket q_{2}\dots\ket q_{N}.
\end{equation}
To get the canonical graph state from this, one can just Fourier transform all qumodes but one, say qumode 1, i.e., do $Q_{j}\mapsto P_{j}$ and $P_{j}\mapsto -Q_{j}$ for $j>1$, which only requires a $\pi/2$ optical phase shift of  qumodes $j>1$ (or a $-\pi/2$ shift of qumode 1 alone) in the laboratory~\cite{Braunstein1998a} and yields the nullifiers
\begin{align}
P'_{+} &= e^{-(N-1)r}\,\(P_{1}-\sum_{j>1}^{N}Q_{j}\) \label{eq:n1}\\
Q_{1j} & = e^{-r}(P_{j}-Q_{1}),\quad \forall j\neq k.\label{eq:n2}
\end{align}
These nullifiers can easily be seen to correspond to the canonical graph of a GHZ state \fig{ghz}, right~\cite{Hein2004}. The difference in connectivity, or valence, of a QC cluster and a GHZ graph is significant: the cluster graph possesses a local structure, i.e., one can define a set of nearest-neighbors, or neighborhood, for each qubit. In contrast, the GHZ state is nonlocal, its only neighborhood is the whole graph, and GHZ states have actually been shown to be ``too entangled'' for one-way QC~\cite{Bacon2009,Bremner2009,Gross2009}.

Note that this graph is actually  equivalent, under local unitaries, to a complete canonical graph~\cite{Hein2004}, which makes the $\mathcal H$ graph and the canonical graphs identical even though this is not mandated by the theorem of \eq{theorem} in this case because {\bf G} is not self-inverse.

\subsection{Finite squeezing}
A natural question then arises of the general relationship --- if any --- between matrices {\bf G} and {\bf V}. This question has been answered in several chronological steps. First, it was shown that multimode squeezing {\em always} produces a cluster state, which yielded a general (though not bijective) relation between  {\bf G} and {\bf V}~\cite{Menicucci2007}. This relation then yielded 
 ${\bf G}={\bf V}$ in the notable particular case  ${\bf G}={\bf G}^{-1}$~\cite{Zaidi2008}. 
 
 Finally, Menicucci, Flammia, and van Loock generalized, in a foundational paper, the CV cluster state formalism to finite squeezing by using the symplectic formalism and complex adjacency matrices ${\bf Z}={\bf V}+i{\bf U}$, where {\bf V} is the  canonical graph adjacency matrix and {\bf U} contains  finite squeezing effects. This yields
 \begin{equation}
\(\vec P - {\bf Z} \vec Q\)\ket{\psi_{G}} = \vec 0\ket{\psi_{G}}.
\end{equation}
where $\ket{\psi_{G}}$ is a finitely squeezed cluster state. Moreover, it was shown that 
 \begin{equation}
{\bf Z} = ie^{-2\kappa\tau\bf G}
\end{equation}
where $\tau$ is the interaction time of the Hamiltonian of \eq H. Taking a self-inverse $\mathcal H$ graph ${\bf G}^{2}=\mathbb1$ yields 
 \begin{equation}
{\bf Z} = i\left[\cosh(2\kappa\tau)\mathbb1-\sinh(2\kappa\tau){\bf G}\right],
\end{equation}
which can be proven, for a bicolorable graph, to be equivalent to the {\bf Z} graph
 \begin{equation}
{\bf Z}' = \tanh(2\kappa\tau){\bf G} + i\,{\rm sech}(2\kappa\tau)\mathbb1 = {\bf V'}+i{\bf U'}.
\end{equation}
thereby confirming the equivalence of ${\bf V'}$ and {\bf G} for self-inverse matrices --- a sufficient but not necessary condition as the GHZ example showed. 

\subsection{Fault tolerance, high-dimensional lattices, and topological qumodes}

Raussendorf formulated fault tolerance for qubit-based one-way quantum computing and showed that using topological error encoding over cluster states yields a remarkable fault tolerance threshold at the 1.4\% error probability level for depolarizing errors, using three-dimensional lattices~\cite{Raussendorf2006}. 

As it turns out, the generation of CV cluster state lattices of higher dimension, or valence, is relatively straightforward over the QOFC: $n$-hypercubic-lattice CV cluster states can be generated by the interference of $n$ identical OPOs, using a fractal construction method~\cite{Wang2014a}. However, it is not clear how the expected needed non-Gaussian nature of the quantum error correcting resource will be factored in, in this case, since $n$-hypercubic-lattice CV cluster states are still Gaussian resources.

Topological properties of qumode states have also been explored theoretically with a proposal to measure entanglement entropy of topological structures such as the toric code~\cite{Demarie2014}.

As we mentioned earlier, a fault tolerant CV cluster-state architecture was also proposed by Menicucci using the GKP encoding~\cite{Menicucci2014ft}. This work determined, for the first time,  the squeezing required to build the GKP resource states for given fault tolerant thresholds. The fact that threshold values for squeezing exist at all was actually the main discovery of Ref.~\cite{Menicucci2014ft}: the existence proof of a CVQC fault tolerance threshold. The corresponding squeezing values for different error rate thresholds (corresponding to different encodings~\cite{Walshe2019}) are given in \tab2. It is  worthwhile at this stage to point out that the current record level of optical squeezing is 15 dB (for a single mode)~\cite{Vahlbruch2016}. 
\begin{table}[htbp]
\caption{Upper-bound~\cite{Menicucci2014ft} squeezing thresholds corresponding to fault tolerance thresholds}
\begin{center}
\begin{tabular}{l|c|c|c}
Desired error rate threshold (encoding-dependent)  & $10^{-2}$ & $10^{-4}$ & $10^{-6}$ \\
\hline
MAXIMUM required squeezing threshold (dB) & 15.6 & 18.7 & 20.5 \\
\end{tabular}
\end{center}
\label{tab:2}
\end{table}%
This result has inspired more theoretical work to now {\em optimize} this threshold to lower values. This result was recently improved by showing that excess technical noise in excess of the reciprocal of the squeezing level --- which is a signature of impurity of the squeezed state --- doesn't affect the QC outcome~\cite{Walshe2019}. Other recent work has shown that fault-tolerant CVQC could be reachable on the order of 10 dB squeezing, using different architectures~\cite{Fukui2018,Fukui2019}. Another avenue deserving of theoretical work is the possibility of non-Gaussian error correcting resources other than GKP states, such as Fock states, which could benefit from the coming of age of photon-number-resolved detection~\cite{Lita2008}. Note that non-Gaussian binomial, a.k.a.\ ``kitten-state,'' error encoding has also been done in the context of superconducting qubits~\cite{Michael2016}. 

The takeaway here is that there are no fundamental limits to fault-tolerant one-way CVQC, even if a great deal of theoretical and experimental work remains to be done.

\section{Experimental realizations of CV multipartite entanglement in the QOFC}

\subsection{Toroidal square lattice proposal}

\begin{figure}[htb]
\centerline{\includegraphics[width=\columnwidth]{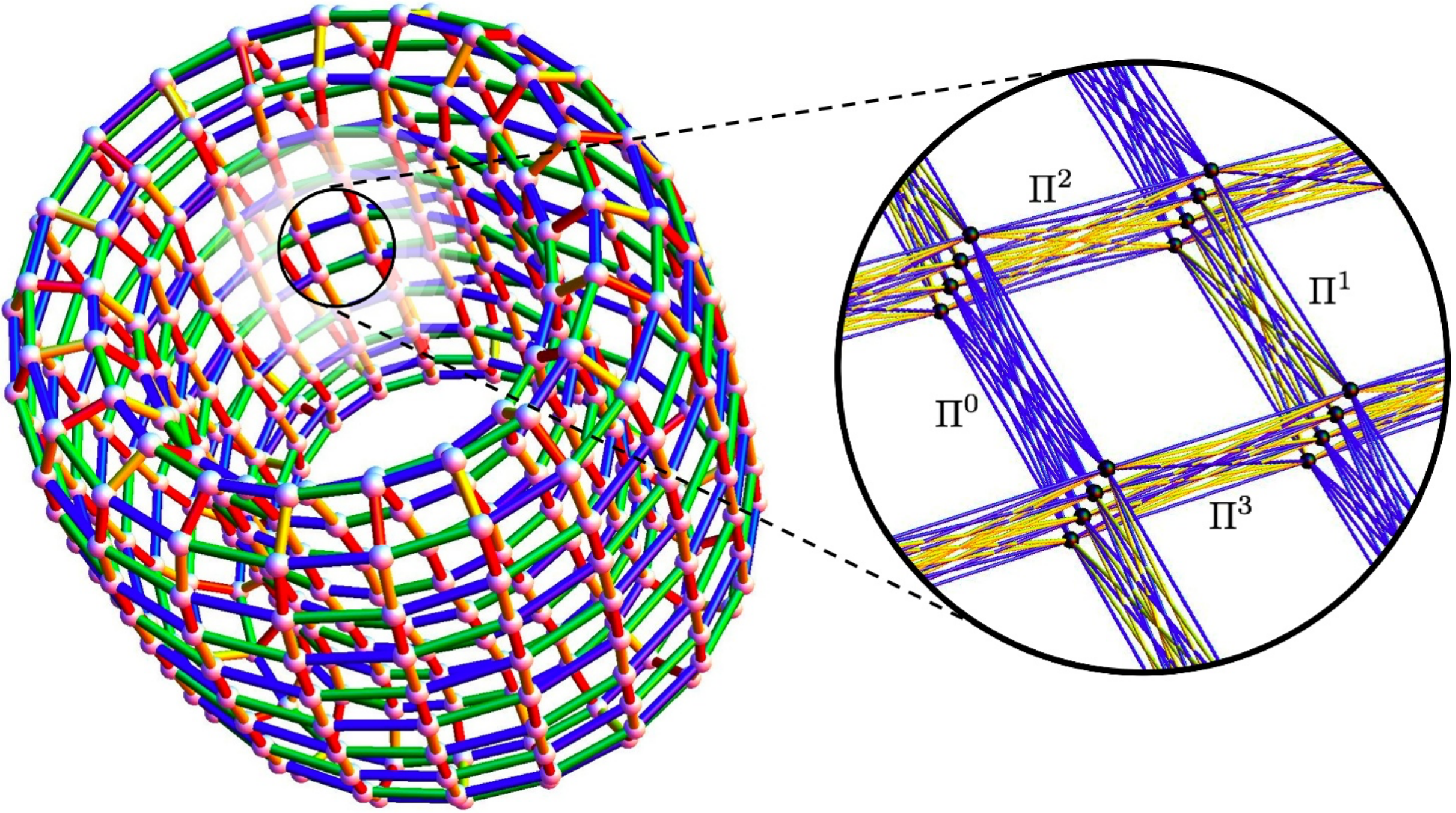}}
\caption{The square lattice cluster state proposed in~\cite{Menicucci2008}. Left, the resulting canonical CV cluster graph, identical to the $\mathcal H$ graph (self-inverse {\bf G}). Right inset, architecture detail: each white vertex is a set of 4 the black vertices which represent individual $\rm TEM_{00}$ cavity qumodes, labeled by 2 frequencies and 2 orthogonal polarizations. The blue and yellow edges denote the ZYY, ZZZ, and YZY/YYZ nonlinear interactions.}
\label{fig:toroid}
\end{figure}

The initial proposal for generating a square-lattice cluster state in a single OPO is depicted in \fig{toroid}~\cite{Menicucci2008,Flammia2009}.
This work overcame a no-go theorem for creating linear-chain and square-lattice cluster states in the QOFC~\cite{Flammia2009}. The solution, as always with impossibility proofs, was to think outside of the box and expand the context of said proof  by adding another degree of freedom, polarization, to the frequency label of qumodes. Conceptually, this allowed the replacement of the regular {\bf G} matrix over QOFC qumodes with a more general matrix whose entries are $2\times2$ polarization blocks. Such a general matrix isn't subject to the aforementioned no-go theorem and can be used to build universal CV cluster states. 

The proposal to implement the needed polarization-block {\bf G} called for a doubly resonant  OPO containing a specially engineered periodically poled $\rm KTiOPO_{4}$ (KTP)  crystal, phasematching the 3 different pump/signal/signal polarization sets ZZZ, ZYY, and YZY/YYZ, all with equal coupling strengths. This crystal was successfully designed and demonstrated experimentally~\cite{Pysher2010}.   

\begin{figure}[htb]
\centerline{\includegraphics[width=\columnwidth]{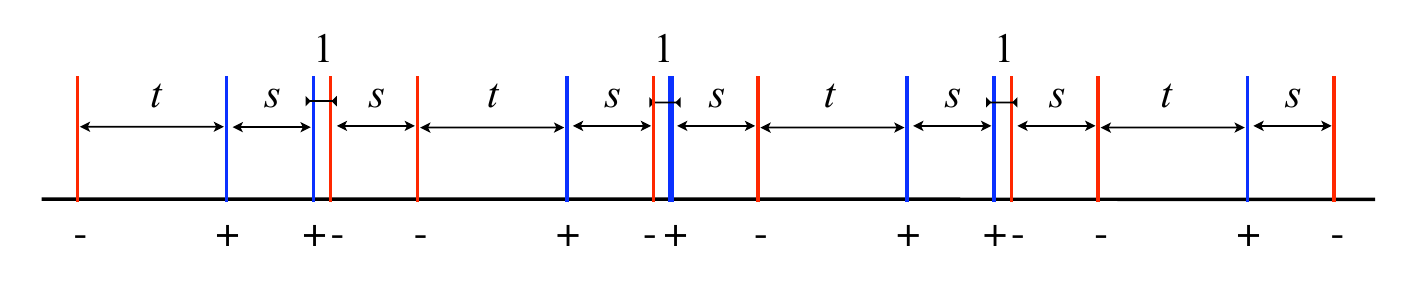}}
\caption{Pump spectrum (scaled by 1/2) of the OPO generating the toroidal cluster state of \fig{toroid}. Parameters $s,t$ are integer multiples of the free spectral range of the OPO cavity, and $\pm$ (blue,red) denote $\pm45^{\circ}$ polarizations.}
\label{fig:pump}
\end{figure}

A slightly inconvenient aspect of this proposal was the fairly complicated 15-mode pump field with orthogonal $\pm45^{\circ}$ polarization components and nontrivial frequency spacings (\fig{pump})
 that would require sophisticated phase modulation techniques to produce, e.g.\ single-sideband modulators~\cite{Izutsu1981} at multiple frequencies. Although possible, the impracticality of this method encouraged the exploration of other avenues for top-down generation of cluster states, starting with smaller sized cluster graphs. This architecture remains, however, remarkably compact and  might still be implementable in the future.

\subsection{Many squares}

\begin{figure}[h!]
\centerline{\includegraphics[width=\columnwidth]{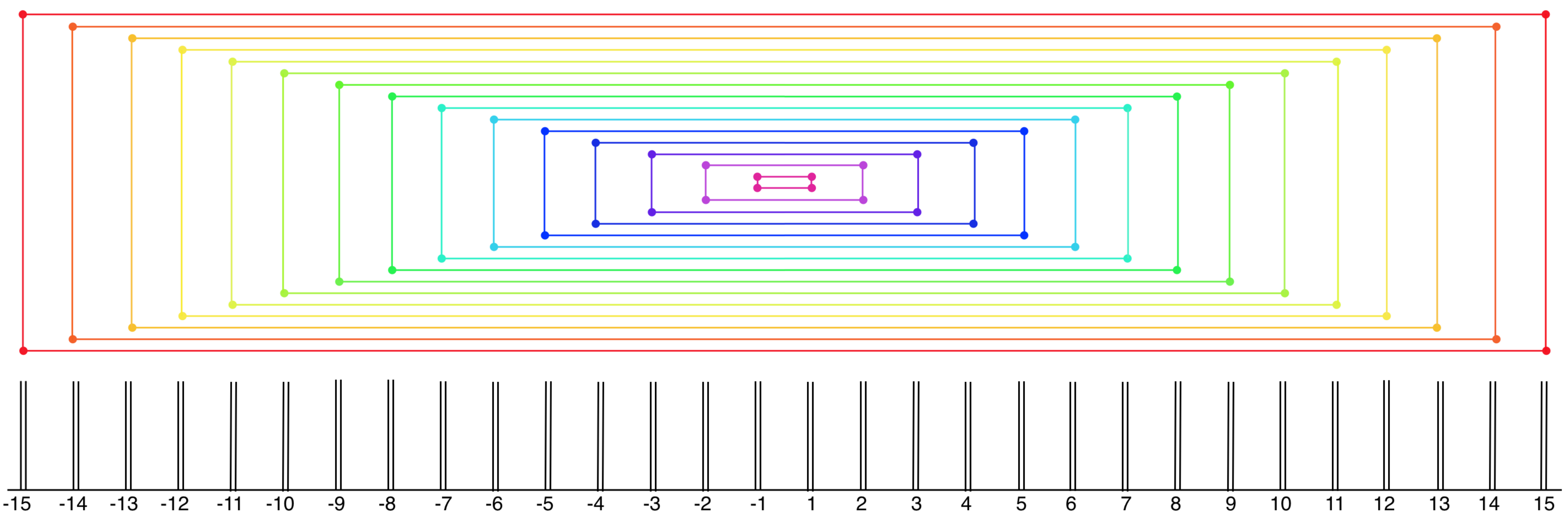}}
\caption{Bottom, the quantum OFC of a single OPO (horizontal axis is frequency, line pairs denote orthogonally polarized modes that are frequency degenerate). Top, canonical CV graph states generated in Ref.~\cite{Pysher2011}.}
\label{fig:msq}
\end{figure}

In 2011, an experiment successfully implemented a 2008 proposal~\cite{Zaidi2008} for creating multiple $2\times 2$ cluster states, \fig{msq}.
Although the cluster state size was small, there was still a novel element of scalability to this work, in the number of copies of the state: 15 copies of the 4-qumode square states were generated simultaneously in the QOFC and verified~\cite{Pysher2011}. This was the first  demonstration of CV cluster state generation over a large scale~\cite{PT2011}. The OPO comprised two KTP  crystals: one PPKTP crystal phasematched the ZZZ and YZY/YZ interactions simultaneously and coupled 2 frequencies and 2 polarizations with a single pump frequency, placing them into ring cluster states; the other crystal was identical to the PPKTP one but unpoled and rotated $90^{\circ}$ with respect to it, this to ensure the crucial requirement of polarization degeneracy of cavity modes at the same frequency.

\subsection{Dual-rail quantum wire}

\begin{figure}[htb]
\centerline{\includegraphics[width=\columnwidth]{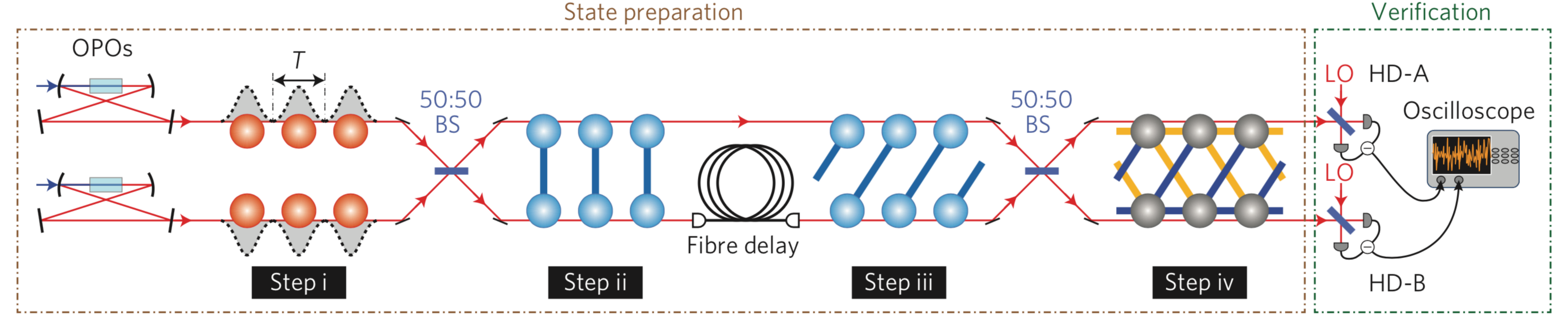}}
\caption{Temporal quantum wire generation, from~\cite{Yokoyama2013}. See text.}
\label{fig:ncm}
\end{figure}

\begin{figure}[htb]
\centerline{\includegraphics[width=.7\columnwidth]{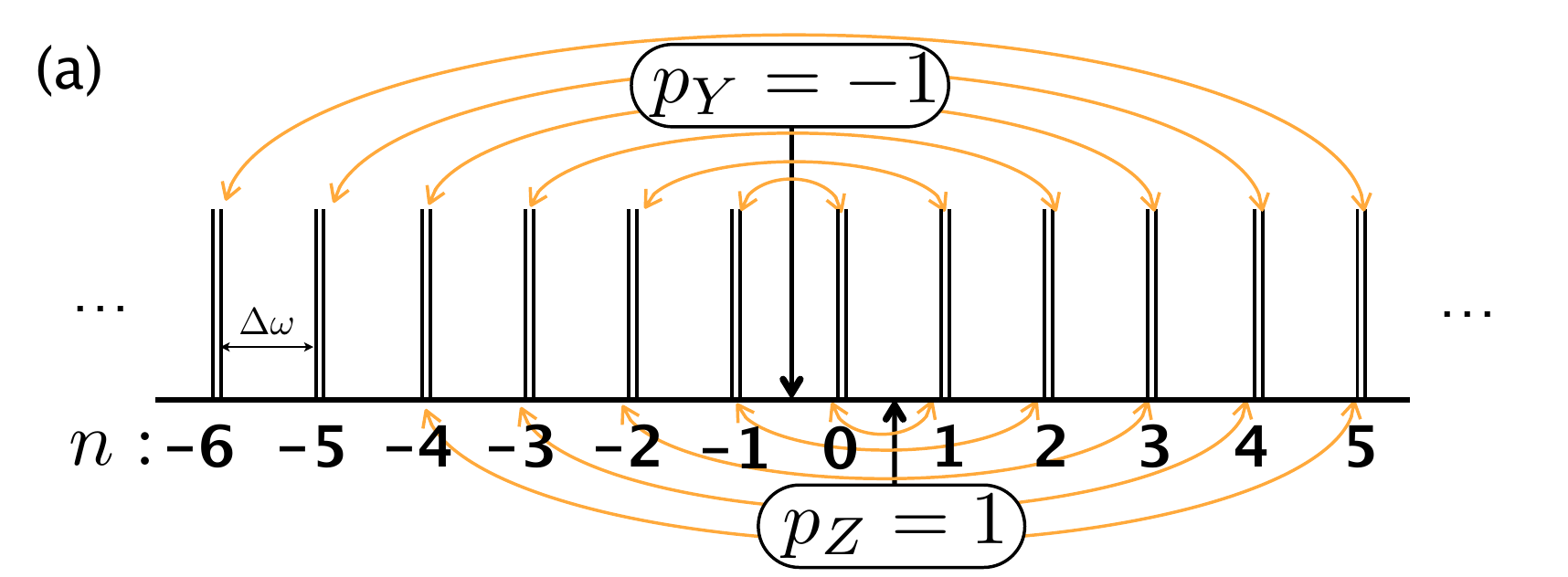}}
\centerline{\includegraphics[width=.7\columnwidth]{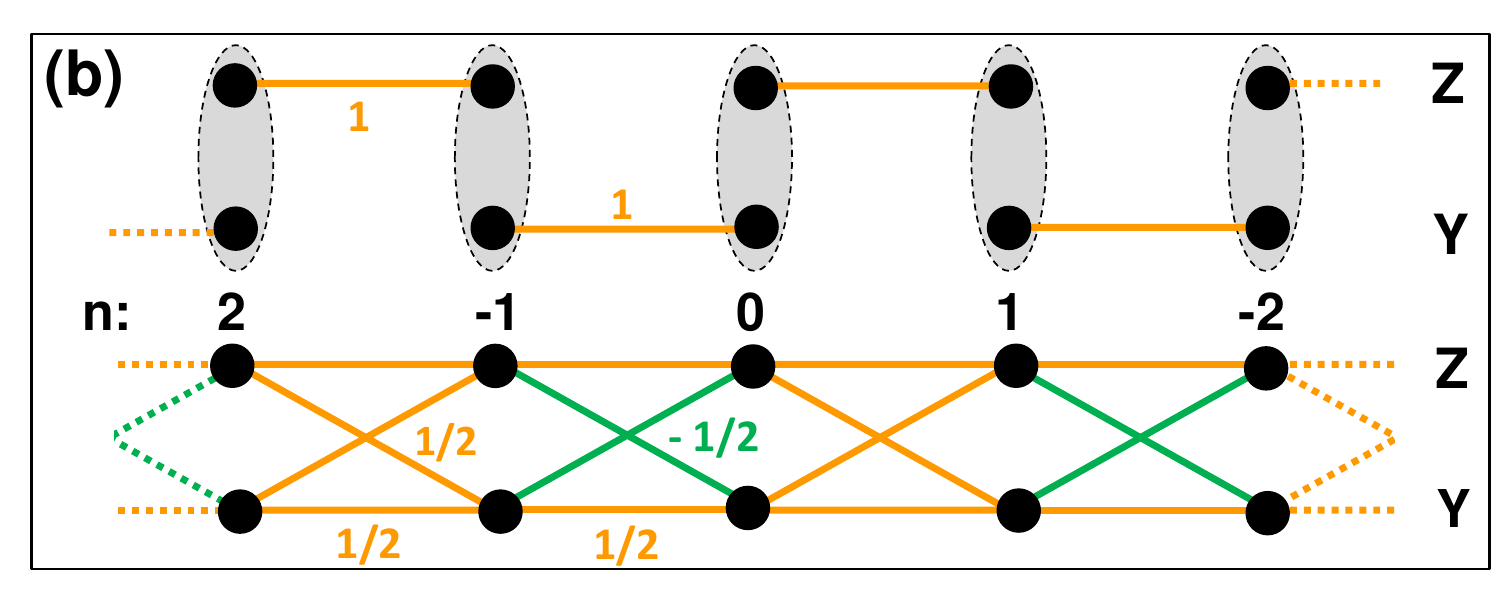}}
\centerline{\includegraphics[width=.7\columnwidth]{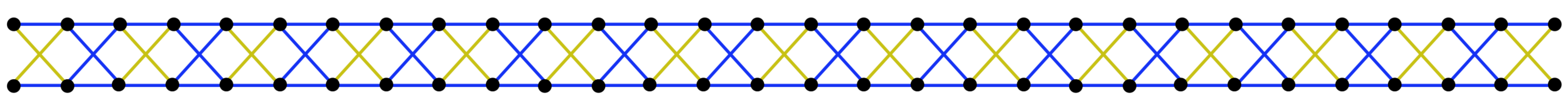}}
\caption{Spectral quantum wire generation, from~\cite{Chen2014}.(a), the initial $\mathcal H$ graph in the QOFC. The arrows mark the pumps' half frequencies. (b), the reordered frequencies make the chain structure appear. The grayed ovals represent balanced beamsplitter interactions. At the bottom is the measured 60-qumode CV cluster state. }
\label{fig:pfi}
\end{figure}

Scalability of the size of the cluster state was finally achieved --- remarkably, while keeping the scalability  feature of number of copies --- by adapting in the frequency domain an initial proposal of Menicucci, Ma, and Ralph for sequential CVQC using time-defined qumodes~\cite{Menicucci2010,Menicucci2011a}. The crux of the idea is to start with TMS states, which we will loosely  call EPR pairs from now on, as the primary building blocks and to ``knit up'' a cluster state chain, or ``quantum wire,'' by entangling qumodes from different pairs. This is described in \fig{ncm} for  the originally proposed temporal approach and in \fig{pfi} for the spectral approach.
In the time-domain, spatially separated EPR pairs are created by interfering two single-mode squeezed states in quadrature to create EPR pairs. Then one qumode of one EPR pair goes through a delay line before interfering with one qumode from the next EPR pair at a balanced beam splitter, resulting in a dual-rail quantum wire structure. This was realized experimentally by Akira Furusawa's group at the University of Tokyo, reaching initial wire lengths of $10^{4}$ qumodes~\cite{Yokoyama2013} and later one million qumodes~\cite{Yoshikawa2016}, accessible sequentially, 2 at a time, see \fig{ncm}. Note this sequential aspect is compatible with QC and has been dubbed the ``Wallace and Gromit approach,'' as explained in Ref.~\cite{Menicucci2010}.  

In the frequency domain version of the original scheme, \fig{pfi}(a), 
all entangled qumodes are generated simultaneously. Rather than straddling two distinct spatial paths and many temporal bins, as in \fig{ncm}, the EPR pairs straddle two orthogonal polarization states  and many frequencies: they are created in two sets, at two orthogonal linear polarizations, and the pairs at one polarization are shifted with respect to the pair at another polarization by frequency shifting the independent pump fields that create each pair set, \fig{pfi}(a). Note that, unlike the delay-line shift in the temporal approach, this frequency shift is a lossless operation. All EPR pairs are emitted in the same cavity mode and are subjected to balanced beam splitting by undergoing a $45^{\circ}$ polarization rotation in a half-wave plate, thereby generating a (slightly different, see edge colors, which denote weight signs) dual-rail quantum wire, \fig{pfi}(b), before impinging on a polarizing beam splitter.

\begin{figure}[htb]
\centerline{\includegraphics[width=\columnwidth]{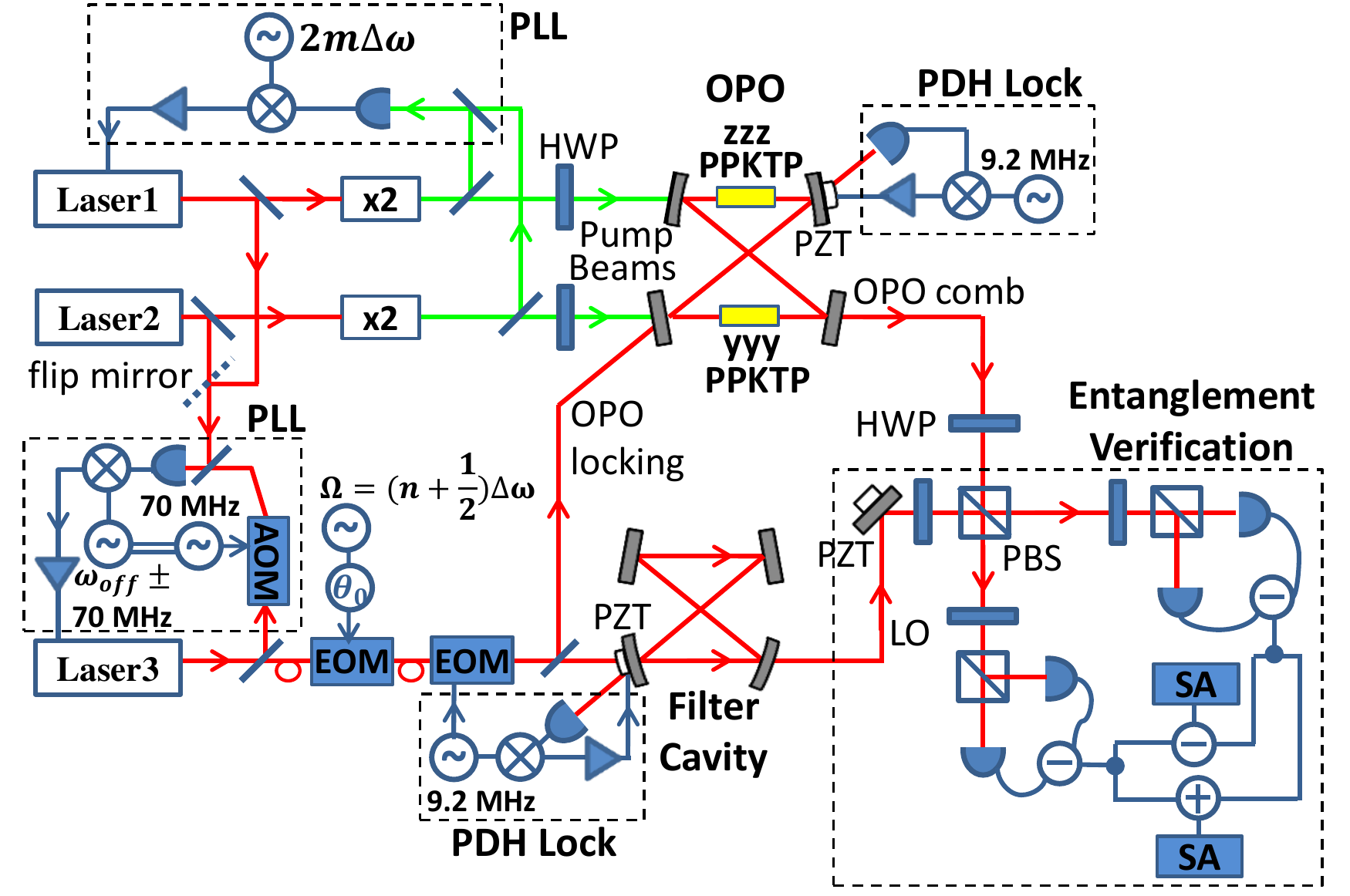}}
\caption{Experimental setup for spectral quantum wire generation. Four servo loops are required to phaselock all CW Nd:YAG lasers together and stabilize the OPO and filter cavities.}
\label{fig:setup}
\end{figure}

\begin{figure}[htb]
\centerline{\includegraphics[width=.8\columnwidth]{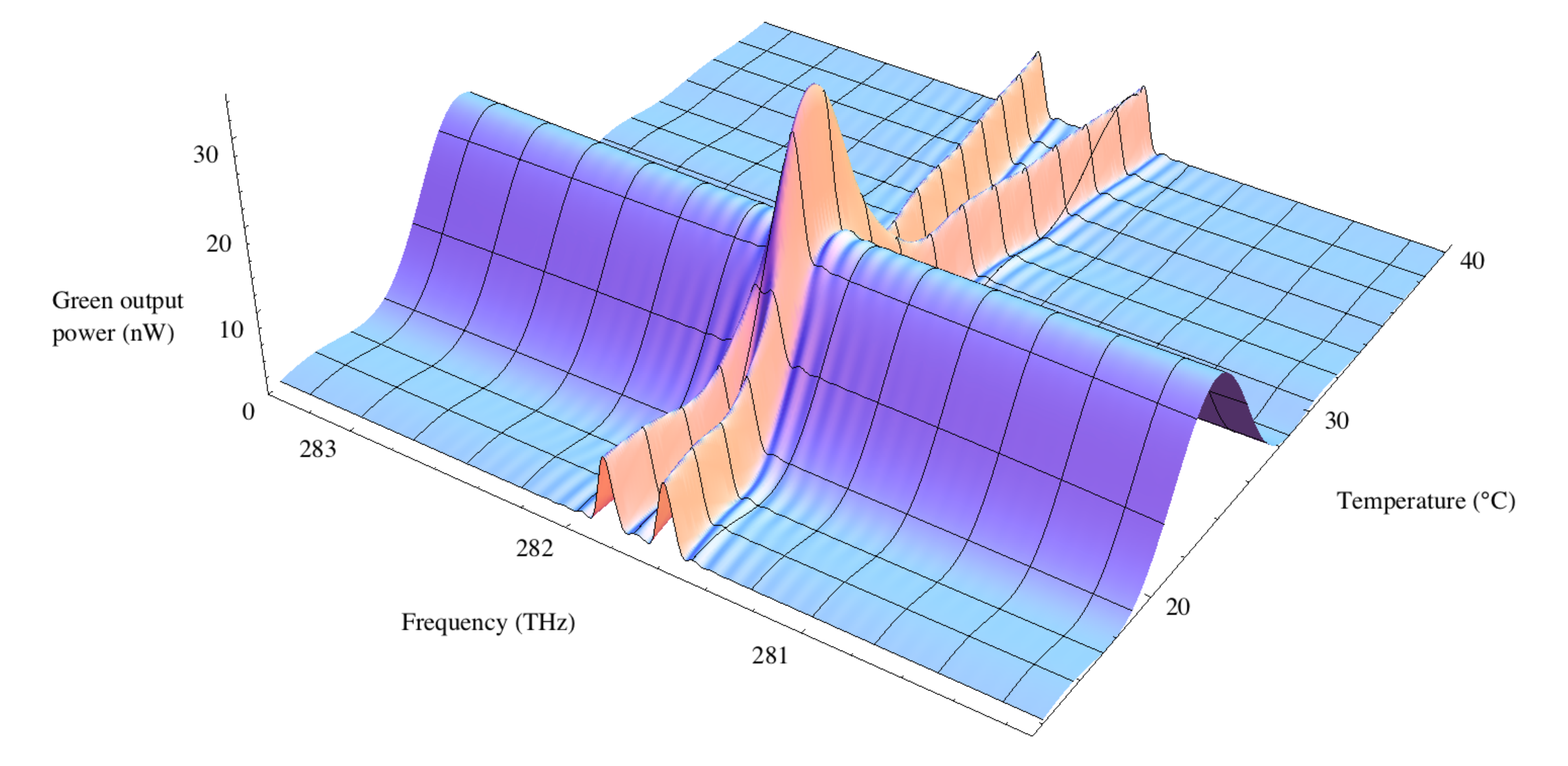}}\centerline{\includegraphics[width=.8\columnwidth]{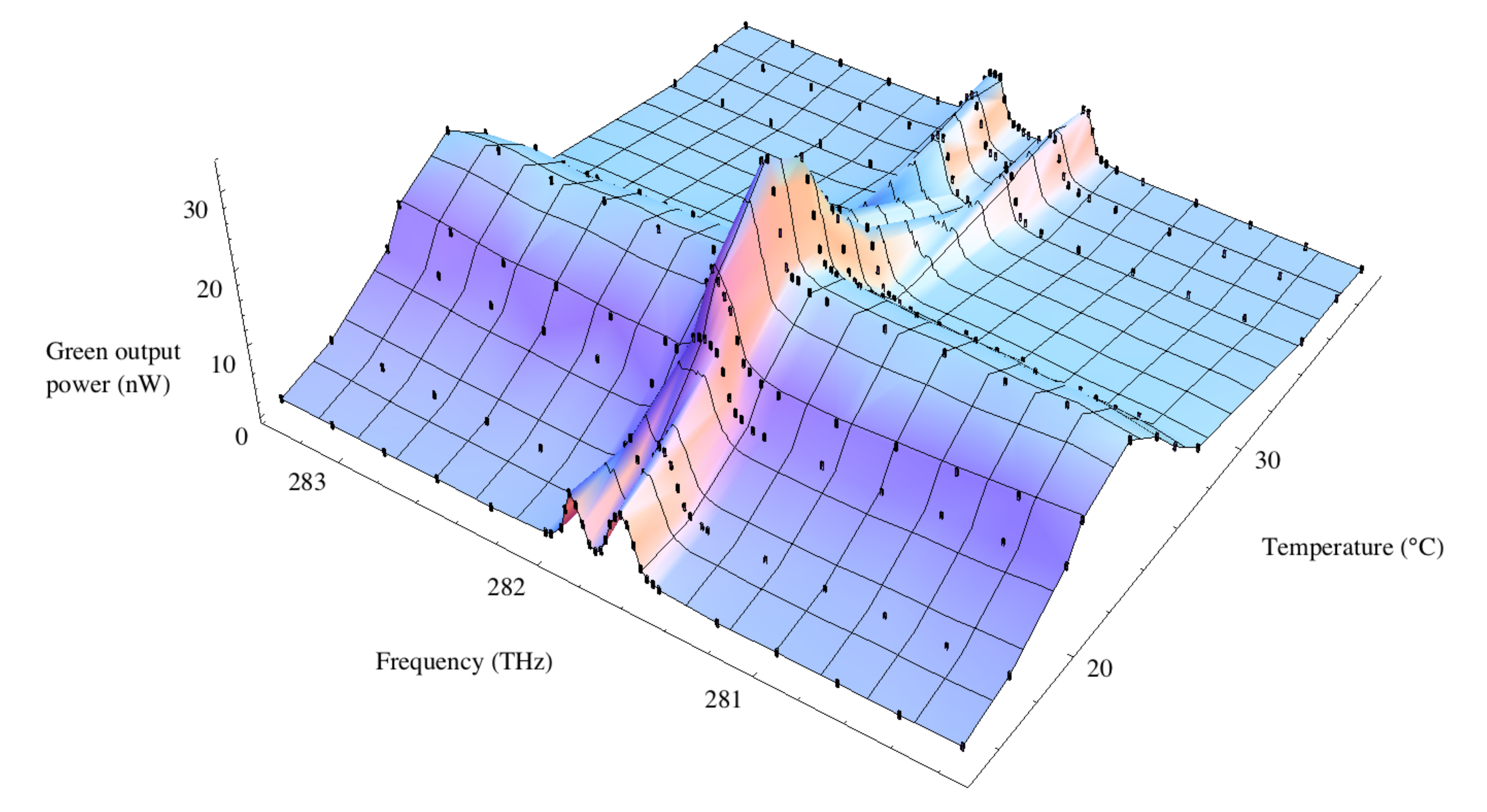}}
\caption{Top, Theoretical phasematching of the SFG process $I_{Z}(\om p)\propto I_{Z}(\om{})I_{Z}(\om p-\om{})$, versus signal frequency $\om{}$ and crystal temperature.  The crossing ridges are due to the more narrowly quasiphasematched SHG interactions. Bottom, Experimental phasematching data. The laser wavelength was scanned from 1058 to 1070 nm, the temperature of the crystal was scanned from 15$^{\circ}$C to 40$^{\circ}$C (11 different temperatures). About 30 data points of different wavelengths were measured at each temperature. The 3D plot was obtained by interpolation (Mathematica) of the data points. The measured SFG bandwidth is 3.178(2) THz, at quasi-constant efficiency, around 23$^{\circ}$C.}
\label{fig:pmbw}
\end{figure}

\Fig{setup} depicts the whole experiment. Three ultrastable CW Nd:YAG lasers (1 kHz emission linewidth) were used to provide tunable pump fields, as well as local oscillator (LO) fields for squeezing detection. 
They also served as frequency references for the two optical cavities in the setup, whose resonance frequencies were locked to them by the Pound-Drever-Hall method~\cite{Drever1983}. All 3 lasers were also phase locked to one another in order to ensure their required precise relative frequency relationships. The filter cavity for the LO field had exactly the same free spectral range (FSR) as the OPO cavity in order to select LO fields corresponding to only 2 OPO QOFC frequencies, at which the balanced homodyne detection setups, one for each polarization, provided a two-tone quantum noise signal. An RF network then reconstructed the nullifier noise, which was found to be squeezed by 3.2(2) dB across the measurement range of 60 modes. This measurement range was determined --- and limited --- by the maximum bandwidth of 15 GHz of the electro-optic modulator used to create the LO fields, which corresponded to 30 QOFC modes (spaced by $0.95$ GHz) at each polarization. This measurement range did not span the whole generation range, which we believed to be at least 3.2 THz from a measurement of the phasematching bandwidth of the OPO PPKTP crystal, as shown in \fig{pmbw}~\cite{Wang2014}.
This measurement used the sum-frequency generation (SFG) of two stable diode lasers, tunable from 1050 nm to 1080 nm, and tuned symmetrically in opposite directions from 1064 nm so as to give a constant 532 nm SFG wavelength, corresponding to that of the OPO pump. In fact, while this measurement showed that the nonlinear interaction inside the OPO, i.e., the squeezing, has constant strength over that 3.2 THz range, corresponding to 6700 OPO qumodes,  it still didn't capture the whole phasematching range as one of the diode lasers ran out of tuning range. The theoretical expectation is closer to 4-5 THz~\cite{Wang2014}, i.e., on the order of $10^4$ modes. It is worth mentioning that this PPKTP ZZZ quasiphasematching bandwidth should increase even more at longer wavelength, to the order of 10 THz for the 775 nm/1550 nm interaction. 

Another limitation will come into play before one runs out of phasematching bandwidth: because of the dispersion of the OPO crystals, the QOFC's FSR will become chirped and qumodes far way from the pump's half-frequency will shift out of OPO resonance. However, this can be remedied by using a slightly spectrally broadened pump field~\cite{Wang2014}. 

\subsection{Dual-rail quantum wire\underline{s} in the QOFC}

\begin{figure}[htb]
\parbox{.65\columnwidth}{\includegraphics[width=.65\columnwidth]{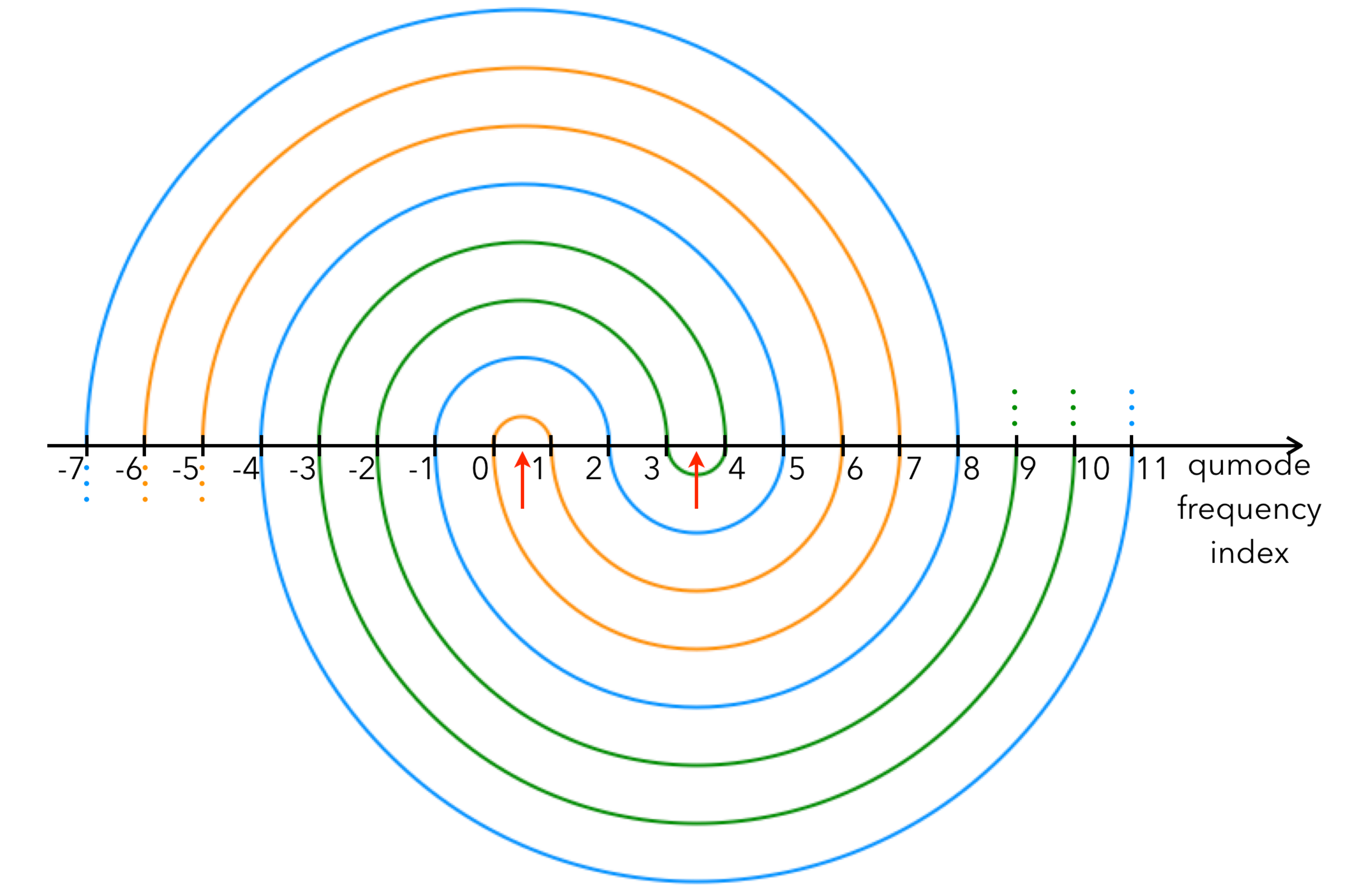}}\parbox{.35\columnwidth}{\includegraphics[width=.35\columnwidth]{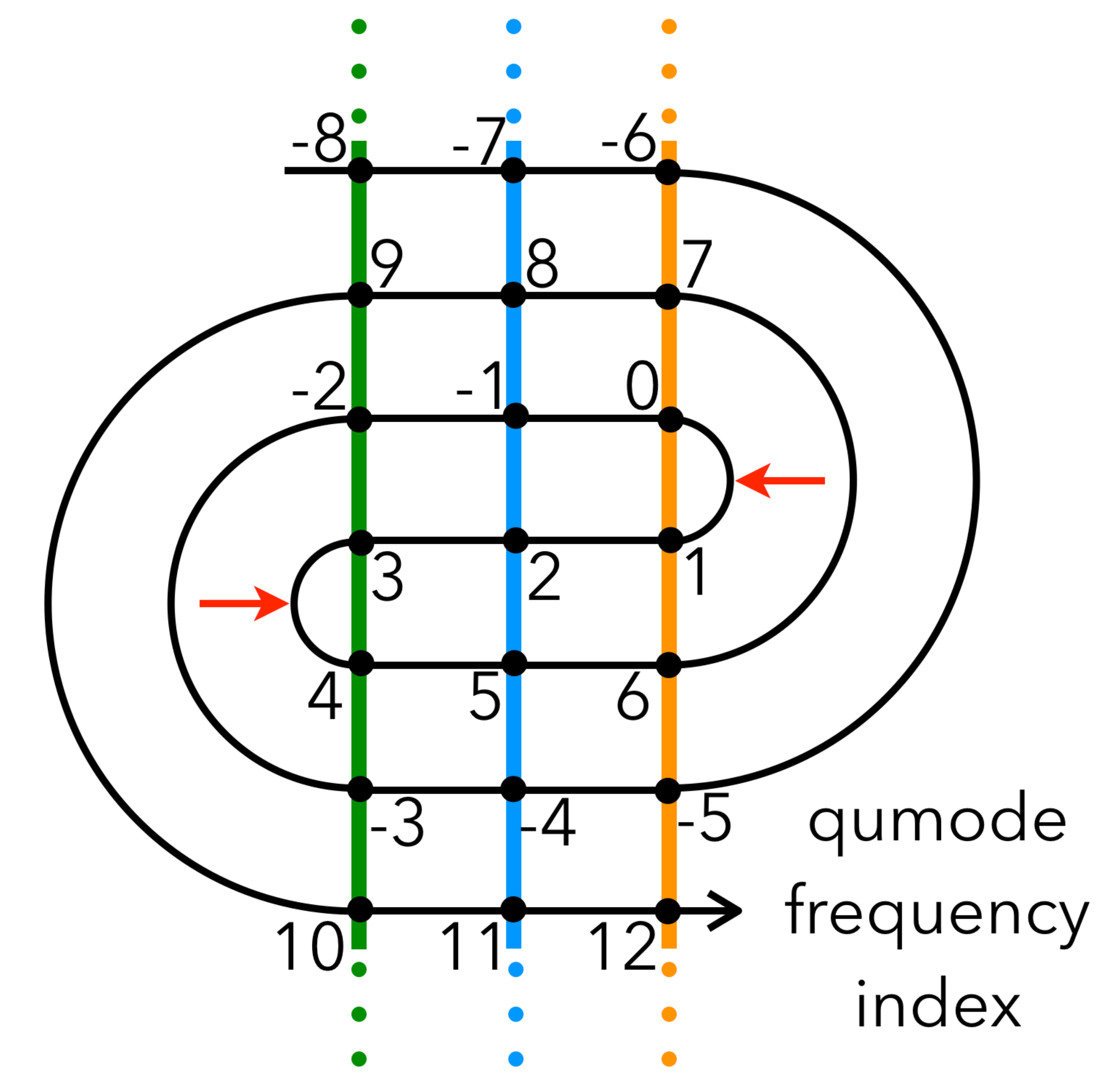}}
\caption{Better living through $\mathcal H$-graph automorphism: scaling of the number of quantum wires by detuning pump fields. The number of quantum wires generated is {\bf 3} in this case, each wire stemming from interactions of one given color (after one additional beam splitter). The red arrows denote the half frequencies of the two pump fields, separated by {\bf 3} FSR.}
\label{fig:auto}
\end{figure}

As was mentioned above, the scalability feature of this scheme doesn't solely apply to the state size, i.e., the number of qumodes per state, it also applies to the number of copies of the state. To see this in the temporal scheme of \fig{ncm}, all one has to do is to consider a temporal delay that is an integer multiple of the mode spacing. Equivalently, in the spectral scheme of \fig{pfi}, one needs detune the pump half-frequencies by an integer multiple of the OPO FSR. This is depicted in \fig{auto}, which presents three equivalent versions of the same $\mathcal H$ graph.
The experimental demonstration of this method yielded two independent quantum wires of 30 (measured) qumodes each generated by a single OPO~\cite{Chen2014}.

An advantage of the spectral implementation is that the large delays required for scaling to large number of wires correspond to large pump detuning, which can be implemented losslessly, in contrast to the temporal implementation if one uses a fiber-based delay line. However, and to be fair, the spectral implementation is ultimately limited by the phasematching bandwidth, i.e., on the order of $10^{4}$ modes in our case, whereas the temporal implementation is only limited by the characteristic stability time of the experiment, which suffers no fundamental limit, being a purely technical issue. 

\subsection{Square and hypercubic cluster states in the QOFC}

The temporal CVQC scheme uses two commensurate delays to ``knit up'' the square lattice cluster state required for universal QC. Experimental implementations of this proposal in the temporal domains were announced very recently and constitute exciting progress, even though scalability was limited by the losses in the temporal delays~\cite{Asavanant2019,Larsen2019}.  

In the spectral domain, the generation of an $N\times N$ square lattice was proposed by interfering two QOFCs, one hosting a single wire (half pump detuning of 1 FSR) with one hosting $N$ independent  wires (half pump detuning of $N$ FSRs)~\cite{Wang2014a}. While this was initially the transposition of the original temporal idea~\cite{Menicucci2011a}, it was discovered in the process that one can expand it to access yet another type of scalability: that of the valence of the cluster graph, from 1D to 2D to hypercubic 4D and above, simply by using 1 QOFC per dimension and generalized interferometers in a fractal procedure~\cite{Wang2014a}. Again, this approach will not suffer from losses in the QOFC shifts but will be capped by the spectral bandwidth of the interaction.

\subsection{Hybrid temporal and spectral approach}

\begin{figure}[htb]
\centerline{\includegraphics[width=\columnwidth]{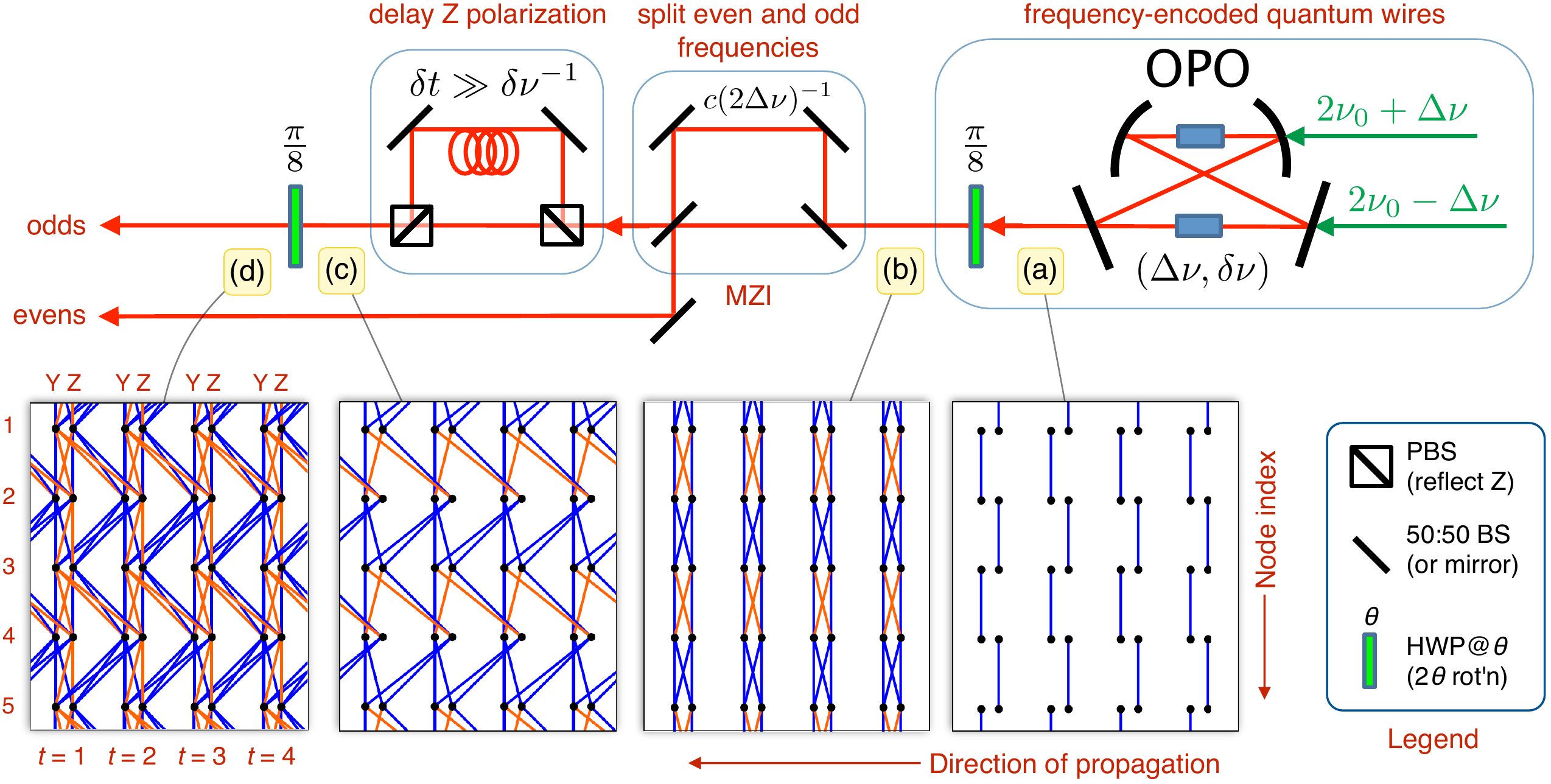}}
\caption{Hybrid spectro-temporal square grid CV cluster state proposal, from Ref.~\cite{Alexander2016b}. See text. The box on the upper right represents the spectral quantum wire experiment of Ref.~\cite{Chen2014} [step (b)]. }
\label{fig:hybrid}
\end{figure}

In an effort to obtain the best out of both worlds, a hybrid approach was proposed in which spectral quantum wires undergo temporal delays and beam splitting to yield square lattice cluster states defined both in the frequency and time domains~\cite{Alexander2016b}. The protocol is depicted in \fig{hybrid}.
It combines the quantum wire generation of Ref.~\cite{Chen2014} with the temporal entanglement of Ref.~\cite{Menicucci2011a}. This creates a square lattice of the ``temporal'' balanced beamsplitter applies to every other frequency mode.  Sorting ``even'' from ``odd'' frequencies in the quantum domain can actually be achieved  fairly easily using a properly unbalanced Mach-Zehnder interferometer~\cite{Huntington2005}. The musical score approximation holds in this case, i.e., the temporal evolution only takes place over time scales much longer than the reciprocal of the linewidth of a qumode. A detailed study of the CVQC protocol in this case showed that such states are, indeed, QC-universal~\cite{Alexander2016b}. 

\subsection{Other implementations of entanglement in the QOFC}

Other degrees of freedom such as transverse spatial modes were used to generate cluster states by Ping Koy Lam's group at the Australian National University~\cite{Armstrong2012}. In the single OPO QOFC, the group of Nicolas Treps at Sorbonne Universit\'e  demonstrated an elegant approach to the generation of QOFC entanglement, by using synchronous pumping, i.e., a mode-locked OFC pump field whose repetition rate is equal to that of the OPO. Using a much broader emission range over which on the order of 10 individually addressable qumodes were defined, they demonstrated qumode-resolved multipartite entanglement~\cite{Roslund2014} and discovered counterintuitive properties of the propagation along a graph of non-Gaussian features induced by photon addition and subtraction~\cite{Walschaers2018}.

\section{Conclusion}

It has been the goal of this review to survey the work done so far on CVQI in the QOFC, as well as with temporally and spatially defined qumodes. 
What's next? As bulk-optics-based approaches continue to explore fundamental concepts, we also look forward for CVQI to translate to integrated platforms and for quantum photonics on chip to become a reality. Much like integrated electronics has been the future of electronic technology, we want to bet on quantum photonics to take this to the next level of scalability and device integration. Quantum technology doesn't yet exist at the level of real-life applications as the challenges, especially that of decoherence, are daunting but it is a worthwhile goal, one that we hope to share with a growing number of researchers.

\section*{References}

\bibliography{/Users/olivierpfister/Dropbox/UVa/Werk/Pfister/Pfister}


\end{document}